\documentclass[aps,amsmath,amssymb,showpacs,showkeys]{revtex4}
\usepackage[dvips]{graphicx,color}
\usepackage{times}
\usepackage{xcolor}
\usepackage[%
  colorlinks=true,
  urlcolor=blue,
  linkcolor=red,
  citecolor=blue
]{hyperref}

\begin{document}
\title{Electromagnetic Quasinormal modes of Dyonic AdS black holes with quasi-topological electromagnetism in a Horndeski gravity theory mimicking EGB gravity at $D \rightarrow 4$ }

\author{Yassine Sekhmani }
\email[Email: ]{sekhmaniyassine@gmail.com}

\affiliation{D\'{e}partement de physique, \'Equipe des Sciences de la mati\`ere et du rayonnement, ESMaR, Facult\'e des Sciences, Universit\'e Mohammed V de Rabat,  Rabat, Morocco}

\author{Dhruba Jyoti Gogoi \footnote{Corresponding author}}
\email[Email: ]{moloydhruba@yahoo.in}

\affiliation{Department of Physics, Dibrugarh University,
Dibrugarh 786004, Assam, India}

%\date{}
\begin{abstract}
 We investigate some properties of a black hole in a Horndeski gravity theory mimicking Einstein-Gauss-Bonnet (EGB) gravity at $D \rightarrow 4$. Borrowing ideas from quasitopological gravities provide a matter source of dyonic fields, in which the black hole solution carries two charges, electric and magnetic, in the context of the EGB gravity. However, due to several limitations of the EGB gravity in $D \rightarrow 4$, we consider a Horndeski gravity theory which can mimic EGB gravity in $D \rightarrow 4$. The essential practice used in this paper is the electromagnetic quasinormal modes process, with the goal of discovering the spectrum of such an electromagnetic perturbation over the black hole spacetime. The Wentzel-Kramer-Brillouin (WKB) approximation is used to achieve the desired results. The study shows that both the charges have similar impacts on the quasinormal modes.
\end{abstract}

%\pacs{04.30.Tv, 04.50.Kd}

\maketitle
\section{Introduction}
Due to their geometric structure, which includes a gravitational singularity region \cite{1a,2a,3a}, black holes are often considered enormous and ambiguous physical objects. Nowadays, many investigations are carried out in an attempt to discover, in some way, the obscure structure of these objects. In this regard, a recent observation was made with the goal of revealing the first image of a black hole, namely the M87* image \cite{4a,5a,6a,7a,8a,9a}, invented by EHT collaboration. Following up on this observation provides a great opportunity to compare a large variety of black holes with respect to size and shape associated with shadow behaviors with those of M87* \cite{41,42,43,44,44a}. Furthermore, another activity serves to discover the hidden mechanism of the black hole in a way that its background geometry is externally perturbed, providing the so-called quasinormal modes. This process is largely interesting and fascinating due to the fact that the black hole is initially perturbed, the scenario where gravitational waves are generated. To see how this works, a distinguished matter field is considered over the spacetime background, allowing the construction of the quasinormal modes spectrum for the black hole. This step is known for a black hole as the ringdown phase. Before an entrance to a relaxed equilibrium, the black hole is in a merger phase, in which there is the emission of gravitational waves with distinctive discrete frequencies known as the quasi-normal modes frequencies, which carry information about the decaying scales and damped oscillation frequencies \cite{10a}. On the other hand, the numerical approach is the most powerful tool to highlight the eigenvalues of quasinormal modes and therefore obtain more information about the intrinsic characterization of the black hole. These numerical approaches are proposed to deal with any framework of gravity \cite{11a,12a} as well as to calculate the quasinormal modes as the continued fraction method \cite{13a}, asymptotic iteration method (AIM) \cite{14a}, Pad\'e approximants \cite{15a}, and the higher-order WKB approximation \cite{16a}. The claim that these numerical approaches are useful for any kind of gravity leads to inspecting the quasinormal modes for several black hole models within the modified gravity, such as the Rastall gravity \cite{17a,18a, 18b, 18c,19a}, EGB gravity \cite{42,20a,21a} and bumblebee gravity \cite{22a}.

Quasi-topological electromagnetism \cite{13} is considered as an alternative way of restoring the dynamic contribution at the stage of the equation of motion.  The reality is that the Maxwell field strength is a topological invariant, as is the Riemann curvature tensor, all of which are $2$-forms,
\begin{equation}
\int tr(R\wedge R), \qquad \int F\wedge F.
\end{equation}
These quantities remain independent of the spacetime metric. To obtain dependency with the metric, it is necessary to take into consideration another purely magnetic field yielding dyonic objects. The starting point is the introduction of supplemented terms within the corresponding Lagrangian, which are associated with  topological invariants. Indeed, in dimension $D=2k$, any $2k$-forms Maxwell field strength is used to build the topological structure, $V_{[2k]}=F_{[2]}\wedge F_{[2]}\wedge ...\wedge F_{[2]}$. In order to carry out this treatment, we consider  the squared norm, combing both electric and magnetic field strengths as follows:
\begin{equation}
U_{[D]}^{(k)}\sim |V_{[2k]}|^{2}\sim V_{[2k]}\wedge \star V_{[2k]}.
\end{equation}
The situation $k=1$ refers to the Maxwell term. Consequently, these invariants remove the non-contribution to the field equations.

The recent finding in the $4\-D$ EGB theory, resulting from the rescale $\alpha\rightarrow\frac{\alpha}{D-4}$ and assuming the limit $(D-4)\rightarrow 0$ \cite{34}. The theory has all of the components of general relativity  and avoids the Lovelock theorem regarding quadratic supplemented terms in $4D$ space-time. The latest activities come up with the essence of $4\-D$ EGB gravity, like shadow behaviors \cite{36,37,38,39} and thermodynamics \cite{d,
014,36,41,42,43,44,44a}. { However, in Ref.s \cite{new01, new02, new03}, it has been argued that choosing $D=4$, leads to an ill defined theory and can have several conceptual issues. To overcome this issue, we follow Ref. \cite{new04} and consider a Horndeski Lagrangian as an alternative of $4\-D$ EGB theory and obtain the black hole solution. This black hole solution is valid for both the Horndeski representation and $4\-D$ EGB theory. Hence, in this paper, our objective is to study the quasinormal modes and a few thermodynamical properties of the black hole solution in a Horndeski gravity theory mimicking EGB gravity at $D \rightarrow 4$.}

The organization of this work is as follows: We construct a black hole solution within Horndeski gravity theory mimicking the $4\-D$ EGB theory with a dyonic matter source in section \ref{sec2}. In section \ref{secnew}, we have briefly investigated the thermodynamical properties of the black hole solution. Section \ref{sec3} aims to investigate the electromagnetic quasinormal modes. The last section \ref{sec4} is devoted to a brief conclusion.

\section{The black hole solution}\label{sec2}
The attempt to have a non-linear electrodynamic field as a source of matter in the context of such a gravity model is governed by the Born-Infeld term \cite{BI} as the main and first model. The pursuit of this model opens the way for new models such as Euler-Heisenberg \cite{EH}, ModMax \cite{MM}, etc. As inspired by these models, quasitopological electromagnetism is considered to be another non-linear electrodynamic field. Certainly, this model borrows ideas from the topological gravity model \cite{TC1,TC2,TC3,TC4}.

An overview of dyonic fields is dedicated to experiencing the situation of a purely electric source \cite{13} with a part of the magnetic field together in a quasi-topological electromagnetism \cite{17,18}. Indeed, We think about the following constraints:
\begin{eqnarray}
\mathcal{F}_{[2k]}&=&F_{[2]}\wedge F_{[2]}\wedge\cdots\wedge F_{[2]}, \,\,\,\, k\leq [D/2] \\
\mathcal{H}_{[pk]}&=&H_{[P]}\wedge H_{[p]}\wedge\cdots\wedge H_{[P]} \,\,\,\,  k\leq [D/p] \\
\mathcal{F}\mathcal{H}_{[2k+pl]}&=& \mathcal{F}_{[2k]}\wedge\mathcal{H}_{[pl]},\,\,\,\,\,\,\,\,\,\,\,\,\,\,\,\,\, \lbrace 2k+p\ell\leq D\rbrace.
\label{Eq1}
\end{eqnarray}
The present step is devoted to constructing the corresponding physical Lagrangian, where the squared norms $\lvert \mathcal{F}_{[2k]}\rvert^2$, $\lvert \mathcal{H}_{[pk]}\rvert^2$ and $\lvert\mathcal{F}\mathcal{H}_{[2k+p\ell]}\rvert^2$ are given in a component notation after using the Hodge star operator,
\begin{eqnarray}
\lvert\mathcal{F}_{[2k]}\rvert^2&\sim&\delta^{\rho_1\cdots\rho_{2k}}_{\sigma_1\cdots\sigma_{2k}}F_{\rho_1\rho_2}F_{\rho_3\rho_4}\cdots F_{\rho_{2k-1}\rho_{2k}}F^{\sigma_1\sigma_2}F^{\sigma_3\sigma_4}\cdots \\
\lvert\mathcal{H}_{[pk]}\rvert^2&\sim&\delta^{\rho_1\cdots\rho_{pk}}_{\sigma_1\cdots\sigma_{pk}}H_{\rho_1\cdots\rho_p}\cdots H_{\cdots\rho_{pk}}H^{\sigma_1\cdots\sigma_p}\cdots H^{\cdots\sigma_{pk}} \\
\lvert\mathcal{F}\mathcal{H}_{2k+p\ell}\rvert^2&\sim&\delta^{\rho_1\cdots\rho_{2k+p\ell}}_{\sigma_1\cdots\sigma_{2k+p\ell}}F_{\rho_1\rho_2}H_{\rho_3\cdots\rho_{p+2}}\cdots F_{\cdots} H_{\cdots\rho_{2k+p\ell}}F^{\sigma_1\sigma_2}H^{\sigma_3\cdots\sigma_{p+2}}\cdots F\cdots H^{\cdots \sigma_{2k+p\ell}},\hspace*{1cm}
\label{Eq2}
\end{eqnarray}
where $\delta^{\rho_1\cdots\rho_{2k}}_{\sigma_1\cdots\sigma_{2k}}$  denotes the rank-$4k$ Kronecker delta. While other quadratic terms can arise differently in a mixed manner between the above quantities, as $\mathcal{F}_{[2k]}\wedge\ast\mathcal{H}_{[p\ell]}$ with $2k=p\ell$ and $k\leq [D/2]$, $\mathcal{F}_{[2k]}\wedge\ast\mathcal{F}\mathcal{H}_{[2q+p\ell]}$ with $p\ell=2(k-q)$ and $k\leq [D/2]$ , as well as $\mathcal{H}_{[pk]}\wedge\ast\mathcal{F}\mathcal{H}_{[2q+p\ell]}$ with $2q=p(k-\ell)$ and $k\leq [D/p]$. In essence, every one of these invariant quantities participates reasonably in the field equations and, as a result, constitutes the matter part of the action.

However, a convenient choice for the gauges fields should be in the following form,
\begin{equation}
F_ {\mu\nu}\sim h'(r)\delta^{x^0x^1}_{\mu\nu},\,\,\,\,\, H_{\rho_1\cdots\rho_p}\sim\delta^{x^2\cdots x^D}_{\rho_1\cdots\rho_p},
\label{Eq3}
\end{equation} 
knowing that  $p=D-2$,  with $F_{\mu\nu}$ and $H_{\rho_1\cdots\rho_p}$ are the electric and magnetic field strength, respectively. From the gauge field structure, it is worth noting that  the only non-vanishing terms are the gauges Kinetic $\lvert \mathcal{F}_{[2]}\rvert^2\sim F_{\mu\nu}F^{\mu\nu}$ and  $\lvert\mathcal{H}_{[p]}\rvert^2$, and an interaction part $\lvert\mathcal{F}\mathcal{H}_{[D]}\rvert^2$ aforementioned. Consequently, we consider a $D$-dimensional action  referred to a gravity sector  in the essence of the EGB theory, and in a part to a matter field labeled by a quasi-topological electromagnetism in the form,
\begin{equation}
S_D\left[g_{\mu\nu},A_\mu,B_{\lambda_1\cdots\lambda_{p-1}}\right]=\frac{1}{16\pi}\int d^Dx\sqrt{-g}\left(R-2\Lambda+\frac{\alpha}{D-4}\mathcal{G}+\mathcal{L}_{QTE}\right),
\label{Eq4}
\end{equation}
where the matter field is represented by the following quasi-topological electromagnetism Lagrangian,
 \begin{equation}
 \mathcal{L}_{QTE}=-\left(\frac{1}{4}F^2+\frac{1}{2p!}H^2+\beta\mathcal{L}_{int}\right)\label{Eq5}
 \end{equation}
with $F^2=F_{\mu\nu}F^{\mu\nu}$ and $H^2=H_{\rho_1\cdots\rho_p}H^{\rho_1\cdots\rho_p}$, and the interaction term is given by
\begin{equation}
\mathcal{L}_{int}=\delta_{\gamma_1\cdots\gamma_D}^{\lambda_1\cdots\lambda_D}F_{\lambda_1\lambda_2}H_{\lambda_3\cdots\lambda_D}F^{\gamma_1\gamma_2}H^{\gamma_3\cdots\gamma_D}.
\label{Eq6}
\end{equation}
It should be noted that the parameter $\beta$ has a mass dimension of $-2$. 
{ However, as we have mentioned earlier, in the $4\-D$ limit the Gauss-Bonnet term i.e. $\frac{\alpha}{D-4}\mathcal{G}$ in the Lagrangian has critical conceptual issues \cite{new01, new02, new03}. To overcome this issue, we at first derive the field equations for $D$ dimensional case using the Gauss-Bonnet term. Then we consider a Horndeski equivalent Lagrangian from Ref. \cite{new04} in the $4\-D$ limit which can produce exactly the same black hole solution as in the case of $4\-D$ EGB theory.  }

According to Lovelock's theory of gravity, the computation must be restricted to second order for the purpose of obtaining the EGB gravity. The case in which Einstein-Hilbert action arises naturally in a part of the action, in addition to the quadratic Gauss-Bonnet term, is given as,
 \begin{equation}
 \mathcal{G}=R^2-4R^{\mu\nu}R_{\mu\nu}+R^{\mu\nu\rho\sigma}R_{\mu\nu\rho\sigma}.
 \label{Eq7}
 \end{equation}
By varying the action with respect to the presence fields, we get the following equation of motions
   \begin{eqnarray}
G_{\mu\nu}+g_{\mu\nu}\Lambda+\alpha L_{\mu\nu}=-\frac{1}{2}F_{\mu\rho}F_{\nu}^{\rho}+\frac{1}{8}g_{\mu\nu}F^2-\frac{1}{4}\mathcal{B}_{\mu\nu}&-&\frac{\beta}{2}g_{\mu\nu}\mathcal{L}_{int}\\
\Delta_\nu F^{\nu\mu}-4\beta\,\delta^{\mu\nu\, \gamma_1\cdots \gamma_p}_{\lambda\cdots \lambda_D} H_{\gamma_1\cdots \gamma_p}\,\Delta_\nu\left(F^{\lambda_1\lambda_2}H^{\gamma_3\cdots \gamma_D}\right)&=&0 \\ 
\Delta H^{\mu \lambda_1\cdots \lambda_{p-1}}+2\alpha\, p!\, \delta^{\mu\nu\rho\, \lambda_1\cdots \lambda_{p-1}}_{\gamma\cdots \gamma_D} F_{\mu\nu}\Delta_\rho\left(F^{\gamma_1\gamma_2}H^{\gamma\cdots \gamma_D}\right)&=&0.
\label{Eq8}
\end{eqnarray}
where $G_{\mu\nu}$ and $L_{\mu\nu}$, respectively, are the Einstein tensor and the Lanczos tensor. They are given by
\begin{eqnarray}
G_{\mu\nu}&=&R_{\mu\nu}-\frac{1}{2}g_{\mu\nu}R, \\
L_{\mu\nu}&=&2\left(R R_{\mu\nu}-2R_{\mu\rho}R_\nu^\rho-  2R^{\rho\lambda}R_{\mu\nu\rho\lambda}+R_\mu^{\rho\lambda\sigma}R_{\nu\rho\lambda\sigma}\right)-\frac{1}{2}g_{\mu\nu}\mathcal{G}.\hspace*{0.6cm}
\label{Eq9}
\end{eqnarray}  
Moreover, $\mathcal{B}_{\mu\nu}$ represents the energy-momentum tensor for the $B_{[p-1]}$ field, which can be written as  
   \begin{equation}
\mathcal{B}_{\mu\nu}=\frac{1}{(p-1)!}H_{\mu \rho_1\cdots \rho_{p-1}}H^{\rho_1\cdots \rho_{p-1}}-\frac{1}{(p!)^2}\delta_{\sigma_1\cdots \sigma_p}^{\rho_1\cdots \rho_p} {^\lambda}(_\mu g_\nu)_\lambda H_{\rho_1\cdots \rho_p}H^{\sigma\cdots \sigma_p}.
\label{Eq10}
\end{equation}
An examination is carried out to determine what exactly can result from the variation of the interaction part of the Lagrangian with respect to the metric space-time. Let us now assume the following:
\begin{equation}
\frac{1}{\sqrt{-g}}\frac{\delta\left(\sqrt{-g}\mathcal{L}_{int}\right)}{\delta g^{\mu\nu}}=X_{\mu\nu}-\frac{1}{2}g_{\mu\nu}\mathcal{L}_{int}.
\label{Eq11}
\end{equation}
   Nevertheless these Lagrangians fulfill the identity
\begin{equation}
\delta^{\rho_1\cdots\rho_D}_{\sigma_1\cdots\sigma_D} F_{[\rho_1\rho_2}H_{\rho_3\cdots \rho_D}F^{\sigma_1\sigma_2}H^{\sigma_3\cdots\sigma_D}g_{\mu]\nu}=-X_{\mu\nu}+g_{\mu\nu}\mathcal{L}_{int}\\, \hspace{1cm}
\label{Eq12}
\end{equation}   
where, this allows us to put the result,
   \begin{equation}
\frac{1}{\sqrt{-g}}\frac{\delta\left(\sqrt{-g}\mathcal{L}_{int}\right)}{\delta g^{\mu\nu}}=\frac{1}{2}g_{\mu\nu}\mathcal{L}_{int}.
\label{Eq13}
\end{equation}
   Of course, this result confirms the validity of the energy-momentum tensor with regards to the interaction term of the Lagrangian. After that, we look for the appropriate solution pertinent to this background. 
   
   we are interested in finding a physical solution to model the structure of dyonic charges within the EGB theory. For that reason, we assume a metric that is static, spherical, and symmetric,
\begin{equation}
ds^2=-f(r)dt^2+\frac{dr^2}{f(r)}+r^2d\Omega^2_{D-2},
\label{Eq14}
\end{equation}
where one has
\begin{equation}
.d\Omega^2_{D-2}=d\theta_1^2+ \sum_{i=2}^{D-2} \prod_{j=1}^{i-1}\sin^2\theta_jd\theta^2_i,	
\label{Eq15}
\end{equation}
   which is the line element of $(D-2)$-dimensional unit sphere. The connection to differential geometry aids in visualizing the local chart $\lbrace x^i\rbrace$ where $i=1,\cdots, p$, which provides a specific metric  $\epsilon_{ij}$ on the manifold  $\Omega_{D-2}$, with determinant $\epsilon$. This $(D-2)$-unit sphere involves certain magnetic objects to wrap spherical $(D-2)$-cycles covered the volume form, $H_{[D-2]}\sim \text{Vol}(\Omega)$, especially
   \begin{equation}
   H_{\rho_1\cdots\rho_p}=q_m\sqrt{\epsilon}\delta_{\rho_1\cdots\rho_p}^{x^1\cdots x^p}.\label{Eq16}
   \end{equation}
   
 The Maxwell field is purely an electric source,
   \begin{equation}
   F_{\mu\nu}=h'(r)\delta_{\mu\nu}^{tr}\label{Eq17}
   \end{equation}
   where the prime denotes the derivative with respect to $r$. These quantities give rise to such electric and magnetic charges as
   \begin{equation}
   q_e\sim\int_{\Omega_\infty}\star F_{[2]},\qquad q_m\sim\int_{\Omega_\infty} H_{[D-2]}.
   \end{equation}
   
  At this stage, making use of the Ansatz in  Maxwell equations leads to find 
\begin{equation}
r^{2p}\left[p h'(r)+r h''(r)\right]- 8 \beta (p!)^2 q_m^2\left[p h'(r)-r h''(r)\right]=0, \, p=D-2
\label{Eq18}
\end{equation}
   which admit a solution in the form,
   \begin{equation}
   h'(r)=\frac{q_e r^p}{r^{2p}+8\beta(p!)^2q_m^2}.
   \label{Eq19}
   \end{equation}
  A fascinating note from this solution showed that the  parameter interaction and magnetic charge have similar behavior in preserving the dyonic structure of the black hole. 

\begin{figure}

\begin{minipage}{.5\linewidth}
\centering
\includegraphics[scale=.4]{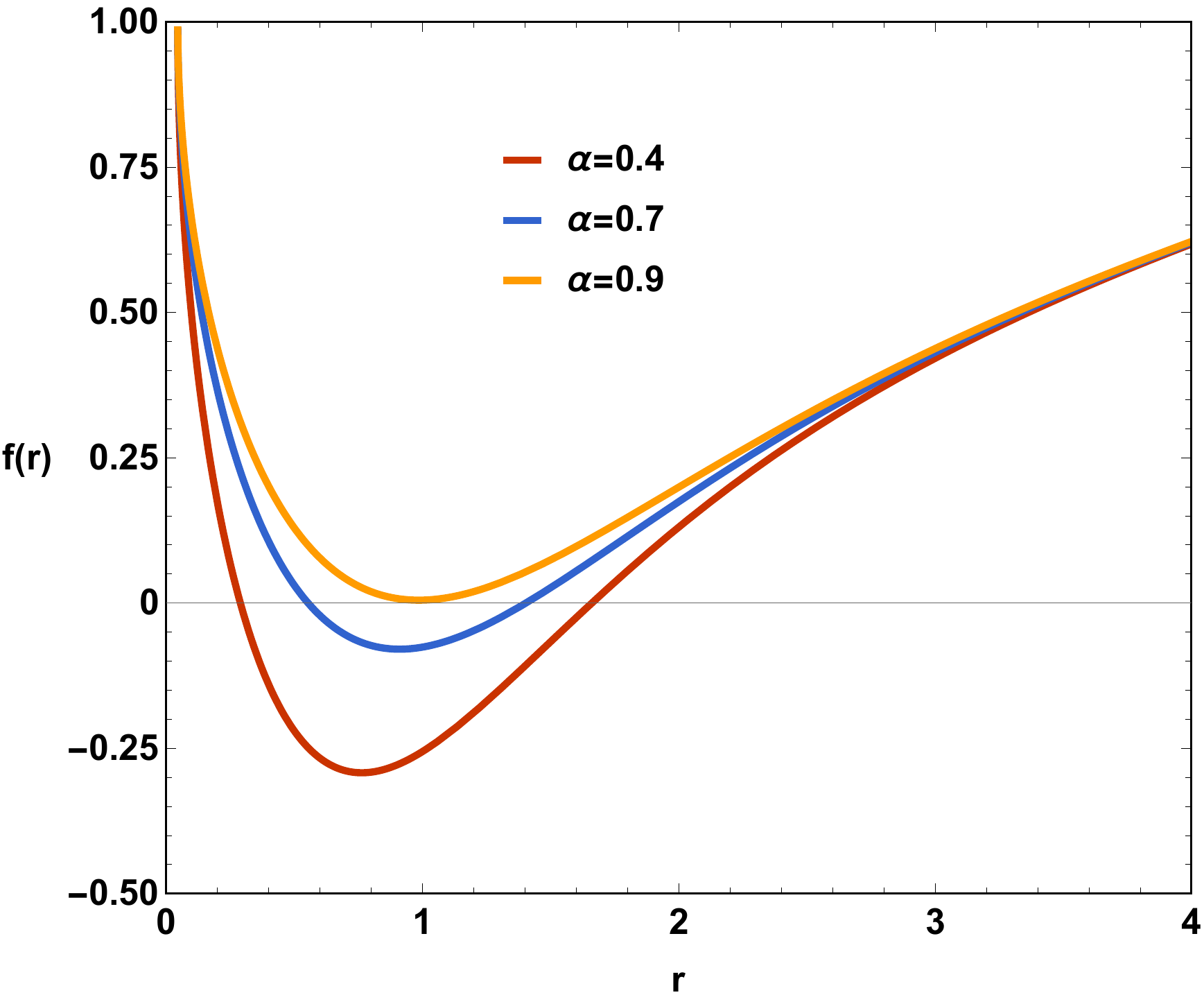}
\end{minipage}%
\begin{minipage}{.5\linewidth}
\centering
\includegraphics[scale=.4]{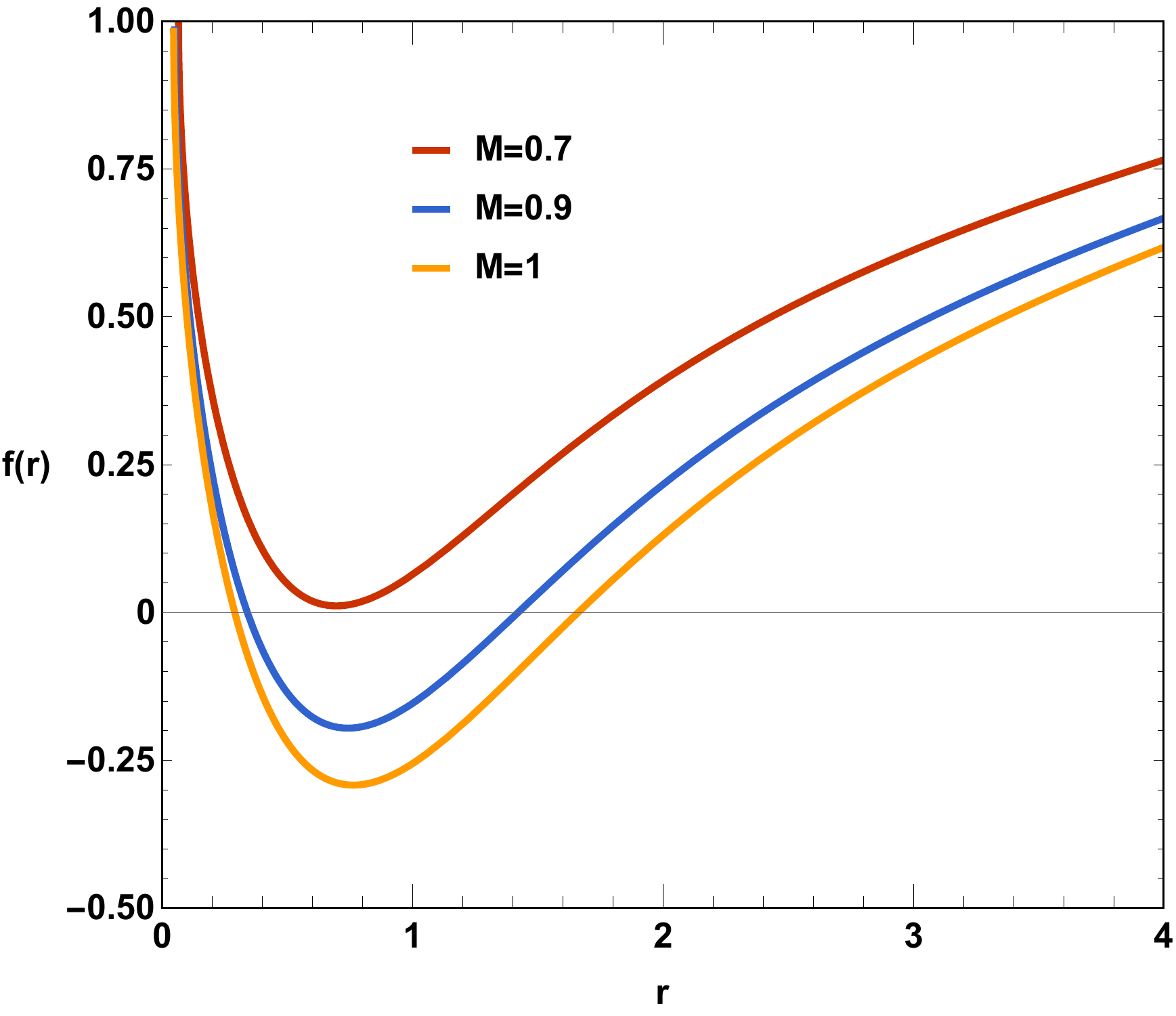}
\end{minipage}\par\medskip

\caption{\label{fig:th} The metric function $f(r)$ for several values of $\alpha$ and $M$ at fixed $q_m=2q_e=0.6$ and $\beta=0.1$. Left panel (a): $M=1$ and right panel (b): $\alpha=0.4$ ($\ell=50$ in both panels)}
\end{figure}

The next step focuses on the definition of the metric function $f(r)$. We begin to evaluate the EGB field equations \ref{Eq8}, reads 
\begin{eqnarray}
T_{t}^t  & = & \frac{D-2}{2}\Bigg(\frac{f'}{r}+\frac{(D-3)f}{r^2}-\frac{D-3}{r^2}\Bigg)-\frac{\alpha(D-2)(D-3)}{2} \Bigg(\frac{2ff'}{r^3}-\frac{2f'}{r^3}+\frac{(D-5)f^2}{r^4}\nonumber\\
&-&\frac{2(D-5)f}{r^4}+ \frac{D-5}{r^4}\bigg)-\frac{(D-1)(D-2)}{2\ell^2},\\
T_t^t&=&T_r^r,
\label{Eq20}
\end{eqnarray}  
where
\begin{equation}
\Lambda=-\frac{(D-1)(D-2)}{2\ell^2}.
\label{Eq21}
\end{equation}
 While the energy momentum tensor is given by
  \begin{equation}
T_t^t=-\frac{1}{4}\left(\frac{q_m^2}{r^{2(D-2)}}+\frac{q_e^2}{r^{2(D-2)}+8\beta q_m^2 \Gamma(d-1)^2}\right).
\label{Eq22}
\end{equation}
{ The case where $D=4$ is particularly interesting, but there is a conceptual issue with this scenario. A recent argument suggests that the EGB part of the action used in Ref. \cite{34} for the 4D case is incorrect, resulting in an ill-defined theory, as demonstrated in Ref.s \cite{new01, new02, new03}. However, a consistent 4D EGB theory can still be constructed, as shown in \cite{new04, new05}. Due to this conceptual issue, the 4D scenario is problematic in the EGB theory. To address this problem, an effective Lagrangian has been proposed as an alternative for the $D\rightarrow 4$ case in EGB theory, as presented in Ref. \cite{new04}. The effective Lagrangian is given by \cite{new04}
\begin{eqnarray}
L_{\rm eff} &=&e^{-\chi}\Big[2(k-\Lambda_0 r^2 - f- r f') +\frac{2}{3}\alpha\phi'\Big(
3 r^2 f^2 \phi '^3+2 r f \phi '^2 \left(-r f'+2 r f \chi '-4 f\right)\notag\\
&&-6 f \phi ' \left(-r f'+2 r f \chi '-f+k\right)-6 (f-k) \left(f'-2 f \chi '\right)\Big)\notag\\
&& + 4\alpha \lambda e^{-2\phi} \Big(r^2 f' \phi '-2 r^2 f \chi ' \phi '-3 r^2 f \phi '^2+r f'+f-k\Big)-6\alpha\lambda^2 r^2 e^{-4\phi}\Big]\,.
\end{eqnarray}
 $(f,\phi)$ form a closed subsystem of equations as
\begin{equation}
f' = f'(f,\phi', \lambda e^{-2\phi})\,,\qquad \phi''=\phi''(f,\phi', \lambda e^{-2\phi})\,,\label{eom}
\end{equation}
and $\chi$ has the following explicit form:
\begin{equation}
\chi' = \dfrac{1}{f} \Big((r\phi'-1)^2 f - \lambda r^2 e^{-2\phi} -k\Big) P(f,\phi', \lambda e^{-2\phi})\,,
\label{chip}
\end{equation}
where $P$ is a complicated expression of rational polynomials of functions $(f,\phi')$ and $\lambda e^{-2\phi}$. When $\lambda=0$, the equations involve $(\phi',\phi'')$ only. To reproduce the 4D EGB scenario, we consider $\chi=0$ \cite{new04}.
}
 The solution can be obtained in the following closed form:
\begin{equation}
f(r)=1+\frac{r^2}{2\alpha}\left\lbrace1\pm\left[1+4\alpha\left(\frac{2M}{r^3}-\frac{1}{\ell^2}-\frac{q_{m}^2+q_e^2 \, _2F_1\left(\frac{1}{4},1;\frac{5}{4};-\frac{32 \beta q_m^2}{r^4}\right)}{4r^4}\right)\right]^{1/2}\right\rbrace.
\label{Eq23}
\end{equation}
where two branches arise naturally within the structure of the solution. In this case, $_{2}F_1$ represents Euler's hypergometric function. Also, once the charges are gone, this causes a Schwarzschild-AdS black hole solution to emerge in the novel $4D$ EGB gravity \cite{34}
\begin{equation}
f(r)=1+\frac{r^2}{2\alpha}\Bigg[1\pm\sqrt{1+4\alpha\left(\frac{2M}{r^3}-\frac{1}{\ell^2}\right)}\Bigg].
\label{Eq24}
\end{equation}
By the way, the solution at a large distance behaves like \begin{equation}
f(r)=1+\frac{r^2}{2\alpha}\left(1\pm\sqrt{1-\frac{4\alpha}{\ell^2}}\right)\pm\frac{8Mr-\left(q_e^2+q_m^2\right)}{4r^2\sqrt{1-\frac{4\alpha}{\ell^2}}}+\mathcal{O}[r^{-4}].
\label{Eq25}
\end{equation}
As a result, the constraint $0<\alpha\leq\ell^2/4$ or $\alpha<0$ imposes the physical validity of the metric function at a large distance. Furthermore, in the vicinity of a small value of $\alpha$, the solution of the plus-sign branch recovers the dyonic RN-AdS solution with a negative gravitational mass and dyonic imaginary charges. Only the minus-sign branch, on the other hand, can reduce to the correct dyonic RN-AdS solution \cite{19}, 
\begin{equation}
f(r)=1-\frac{2M}{r}+\frac{q_e^2+q_m^2}{4r^2}+\frac{r^2}{\ell^2}+\mathcal{O}[\alpha].
\label{Eq26}
\end{equation}
Henceforth, we consider only the negative branch in the rest of this paper. { In Fig. \ref{fig:th}, we have shown the variation of the metric function for different values of $\alpha$ and $M$. One can see that with an increase in $\alpha$ or decrease in $M$, naked singularity may be present.}

\section{Thermodynamical Properties of the black hole}\label{secnew}
{ In this part, we looked into the thermodynamic aspect of our black hole solution. We obtain the thermodynamic quantities, namely, the mass, temperature, entropy, and specific heat. The black hole mass can be found from the metric function solution on the horizon as 
\begin{equation}
M=\frac{q_e^2 \, _2F_1\left(\frac{1}{4},1;\frac{5}{4};-\frac{32 \beta\, q_m^2}{r_+^4}\right)+4 \left(\alpha
   +\frac{r_+^4}{l^2}+r_+^2\right)+q_m^2}{8 r_+}.
\label{Eq27HC}
\end{equation}
The expression of the associated black hole mass encompasses, at certain limits, some previously obtained black hole masses. The case of $\alpha=0$ helps in recovering the defined mass for the dyonic AdS black hole with quasitopological electromagnetism and, together with $q_e =q_m =0$, yields the Schwarzschild-AdS black hole mass.

The Hawking temperature, as a thermodynamic quantity, can be evaluated from the following expression:
\begin{equation}
T=\frac{1}{2\pi}\sqrt{-\frac{1}{2}\partial_\mu\xi_\nu\,\partial^\mu\xi^\nu}=\frac{1}{4\pi}f'(r_+),
\label{Eq28HC}
\end{equation}
from which the associated Hawking temperature is found to be
\begin{equation}
T=\frac{1}{4\pi r_+}\Biggl(\frac{12r_+^8-\ell^2 r_+^4\bigl(q_e^2+q_m^2-4r_+^2+4\alpha\bigr)-32\beta\,q_m^2\bigl(-12r_+^4+\ell^2(q_m^2-4r_+^2+4\alpha)\bigr)}{4\ell^2\,r_+\bigl(32\beta\,q_m^2+r_+^4\bigr)\bigl(r_+^2+2\alpha\bigr)}\Biggr).
\label{Eq29HC}
\end{equation}

\begin{figure}[tbp]
\begin{center}
\includegraphics[width=0.4\linewidth]{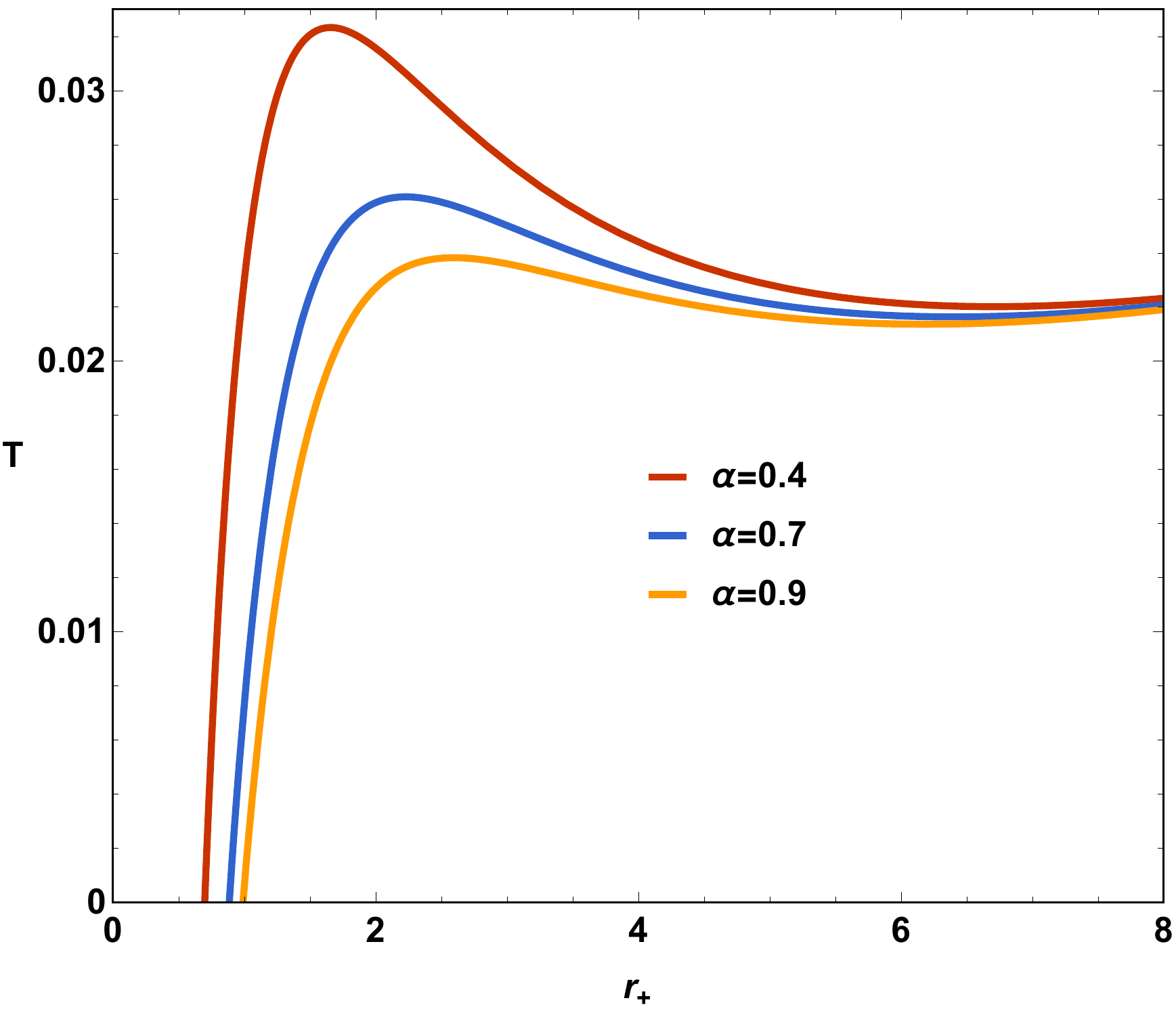}
\end{center}
\caption{\label{fig:T} Hawking temperature $T$ with respect to $r_+$ for several values of  $\alpha$  at fixed value of charges  $q_m=2q_e=0.6$ and $\beta=0.1$ (and $\ell=50$).}
\end{figure}

\begin{figure}[h!]%
 \centering
{\includegraphics[width=0.38\textwidth]{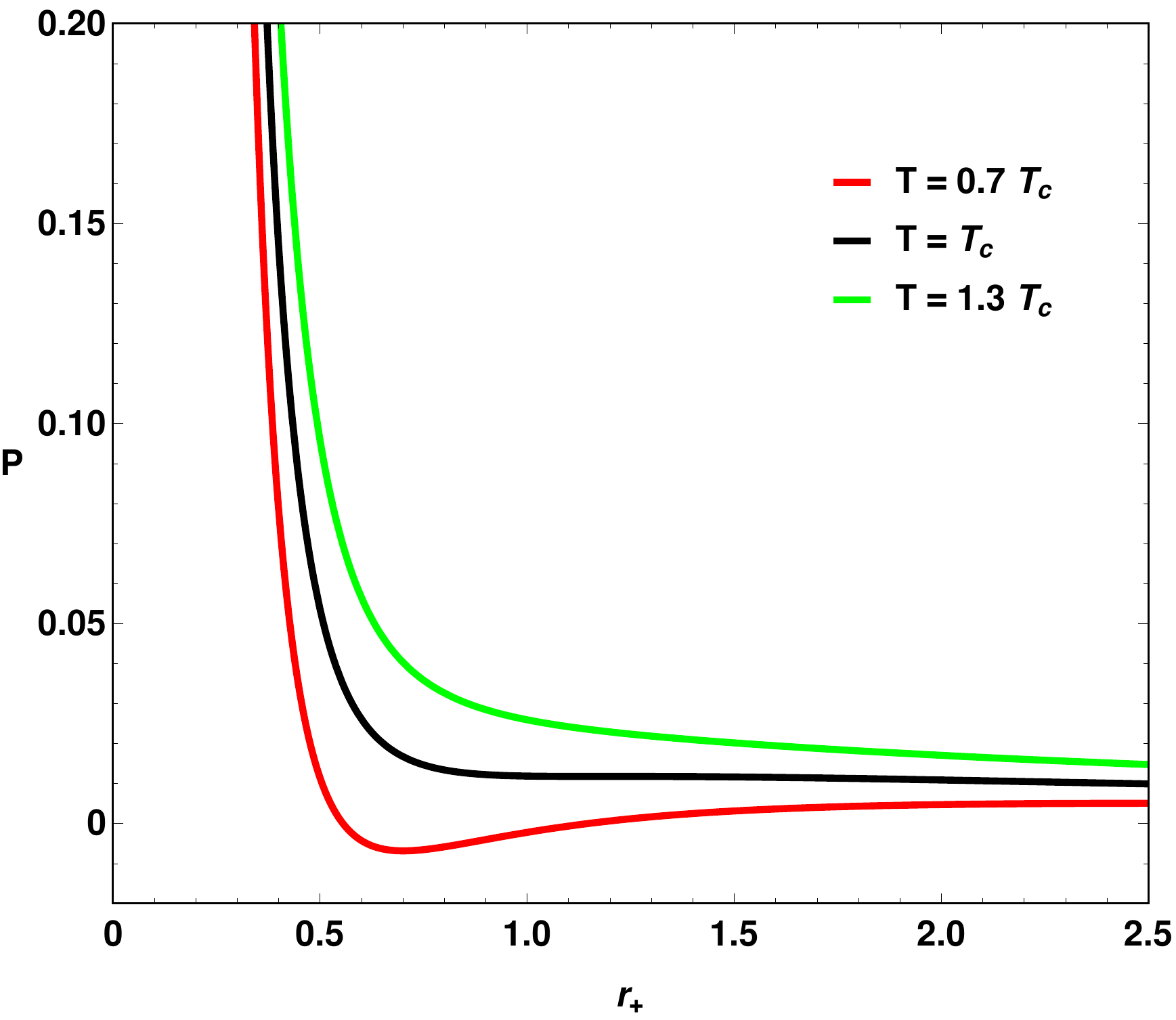}}%

 \caption{Pressure $P$ vs. $r_+$ with $\alpha = 0.1, \beta = 0.1, q_m = 0.2$ and $q_e = 0.2$. The corresponding critical temperature $T_c = 0.078176$ with $r_c = 1.18417$.}%
 \label{HC}%
\end{figure}

\begin{figure}[h!]%
 \centering
{\includegraphics[width=0.38\textwidth]{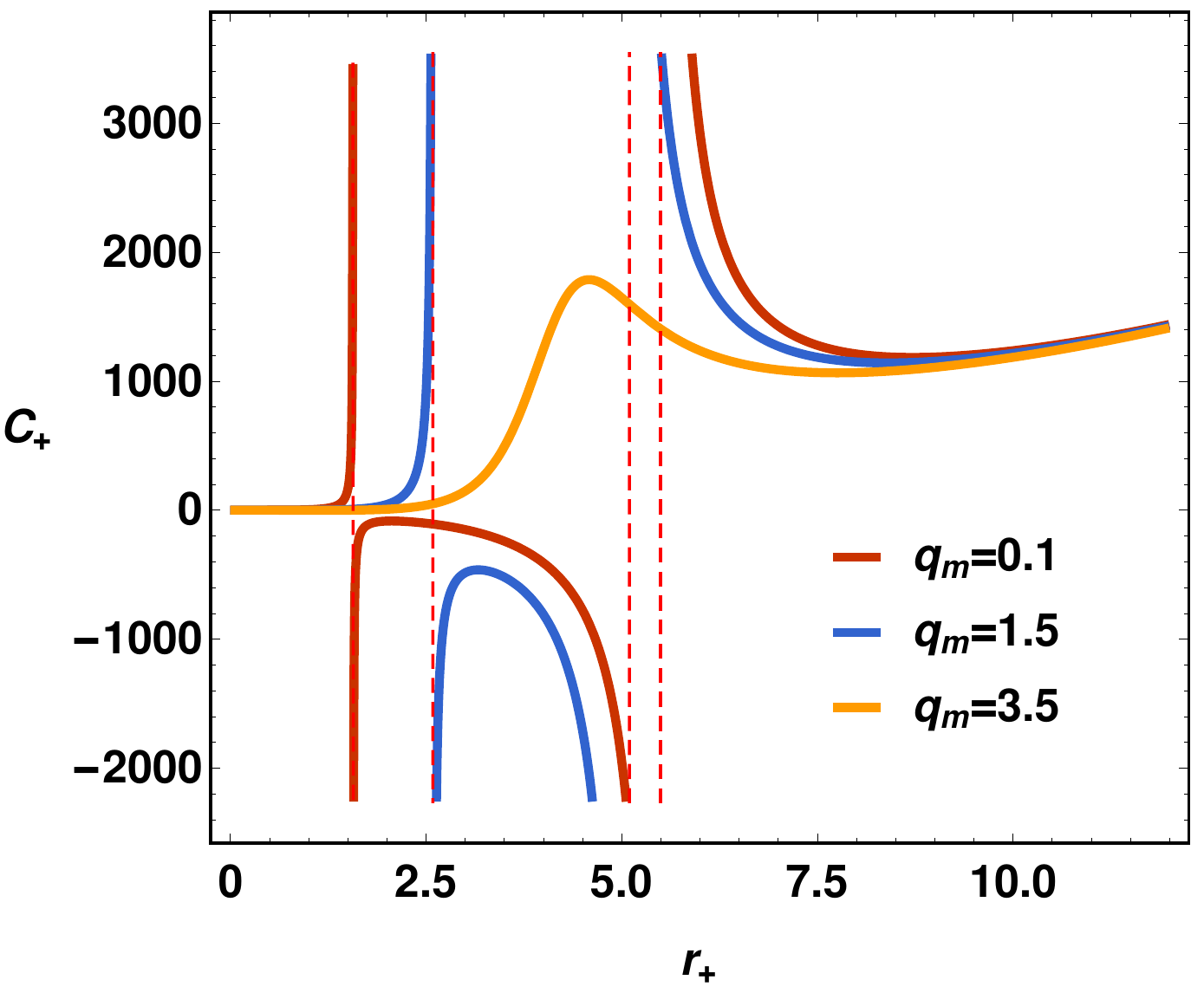}}%
 {\includegraphics[width=0.38\textwidth]{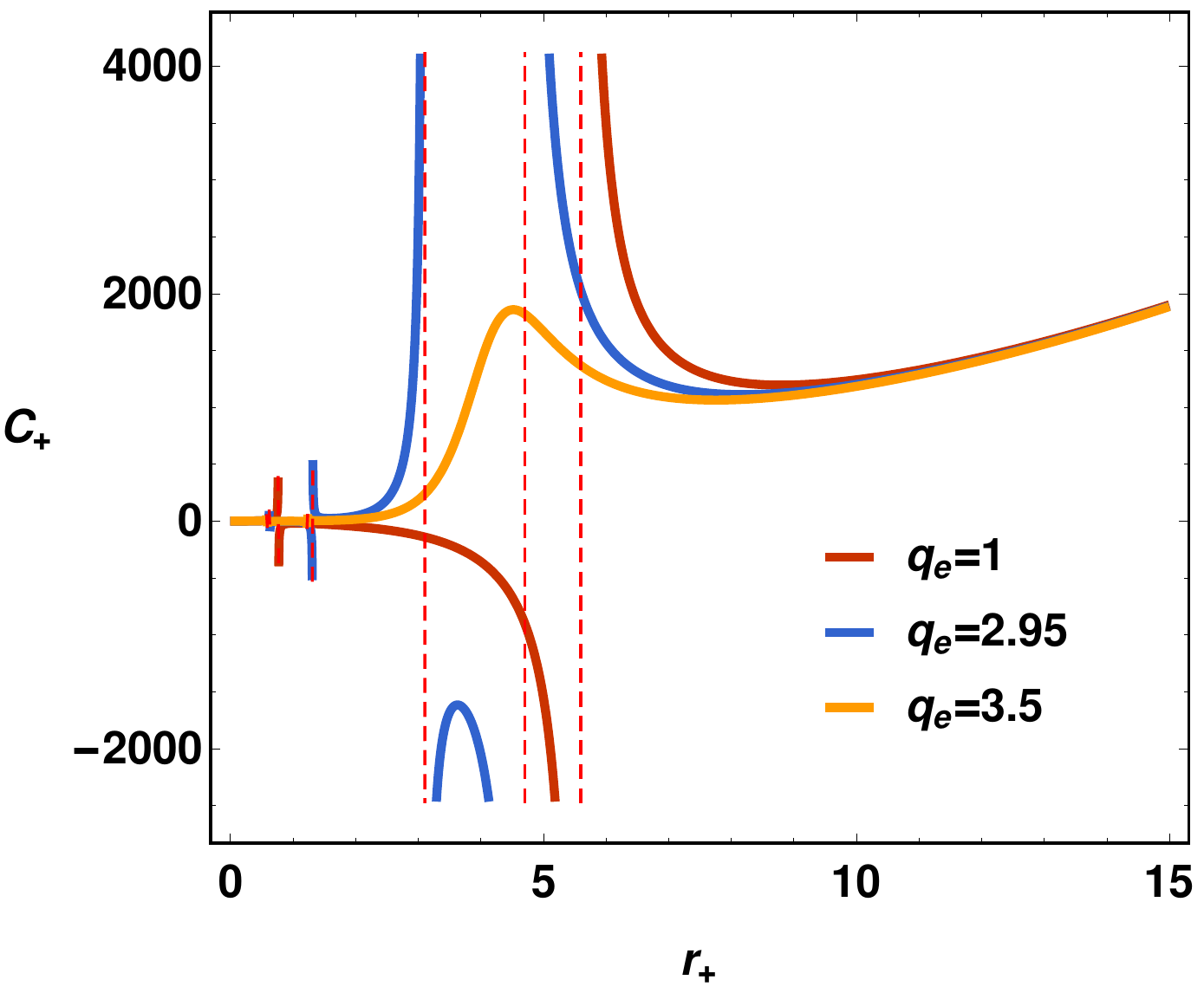}}\\
 {\includegraphics[width=0.38\textwidth]{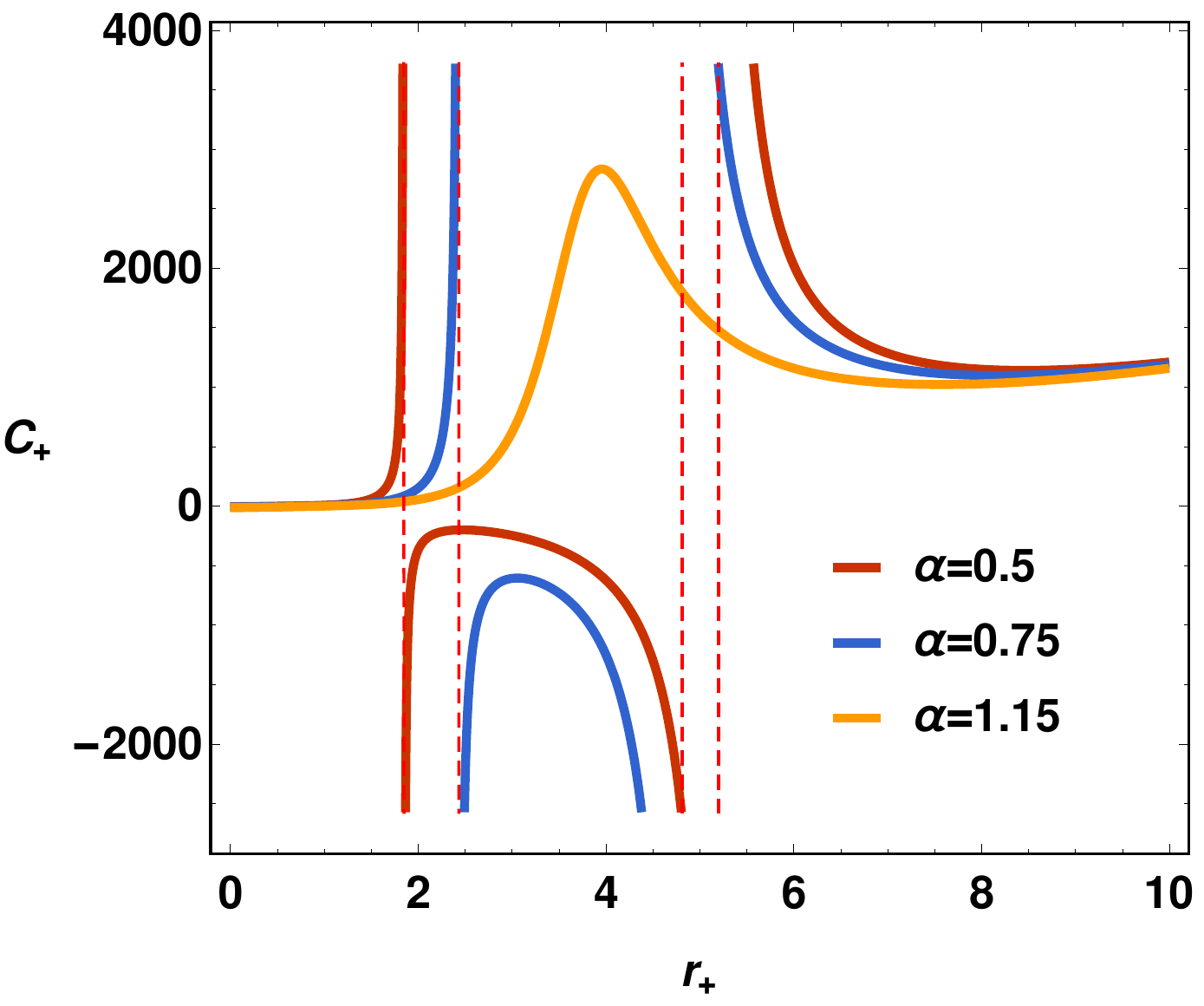}}%
 \caption{Heat capacity behavior against the horizon radius with $M=1, \beta=1$ and $\ell = 10$. One the first panel, we have used $q_e = 0.3$ and $\alpha=0.1$, on the second panel $q_m=0.3$ and $\alpha=0.1$ and on the third panel $q_m=0.3$ and $q_e=0.3$.}%
 \label{HC2}%
\end{figure}

The situation $\alpha \rightarrow 0$ entails a change at the level of the corresponding Hawking temperature, resulting in a $4\-D$ dyonic AdS black hole with quasitopological electromagnetism \cite{17}. Fig.\ref{fig:T} depicts temperature $T$ in terms of horizon radius $r_+$ for various $\alpha$ values.

We define pressure associated with the black hole as
\begin{equation}
P=\frac{3}{4 \pi  \ell^2}.
\end{equation}
Using this relation and $f(r_+)=0$, we can find pressure in terms of the horizon radius as
\begin{equation}
P = -\frac{3 \left(q_e^2 \, _2F_1\left(\frac{1}{4},1;\frac{5}{4};-\frac{32 \beta  q_m^2}{r_+^4}\right)+q_m^2+4 \left(\alpha -2 M r_++r_+^2\right)\right)}{32 \pi  r_+^4}.
\end{equation}
In terms of Hawking temperature $T$,
\begin{equation}
P = \frac{\frac{q_e^2}{32 \beta  q_m^2+r_+^4}+\frac{q_m^2+4 \left(\alpha +r_+ \left(r_+ \left(4 \pi  r_+ T-1\right)+8 \pi  \alpha  T\right)\right)}{r_+^4}}{32 \pi }.
\end{equation}
The thermodynamic volume can be written as \cite{new2023}
\begin{equation}
V = \frac{4 \pi  r_+^3}{3}.
\end{equation}

Hence, the critical point $r_c$ can be found by using $\partial P / \partial r_+ = 0 = \partial^2 P / \partial r_+^2$, which satisfies the following relation:
\begin{align}
r_+^8 \Big[6 \alpha  \left(7 q_e^2-4 \alpha \right)+r_+^2 \left(5 q_e^2-24 \alpha \right)+2 r_+^4\Big]=&\frac{8 r_+^{12} q_e^2 \left(6 \alpha +r_+^2\right)}{32 \beta  q_m^2+r_+^4}+ r_+^4 q_m^2 \Big[1536 \alpha ^2 \beta +2 r_+^4 (3 \alpha -64 \beta )+\notag \\ &1536 \alpha  \beta  r_+^2+3 r_+^6\Big]+64 \beta  q_m^4 \Big[384 \alpha ^2 \beta +r_+^4 (6 \alpha -32 \beta )+ \notag \\ &384 \alpha  \beta  r_+^2+3 r_+^6\Big]+3072 \beta ^2 q_m^6 \left(2 \alpha +r_+^2\right) \Big|_{r_+ = r_c}.
\end{align}
For $\alpha = 0.1, \beta = 0.1, q_m = 0.2$ and $q_e = 0.2$, one can have $r_c = 1.18417.$ The corresponding critical temperature $T_c = 0.078176$. For these values, we have plotted $P$ vs. $r_+$ in Fig. \ref{HC}. For the green curve, we have $T> T_c$, which corresponds to ``ideal" gas phase behaviour. On the other hand, for the red curve, $T<T_c$ and in this scenario, the Van der Waals like small/large black hole phase transition will appear.

It is beneficial to evaluate and carry out a nice study for the thermodynamics aspect, remembering the first law of thermodynamics in a way as \cite{37}
\begin{equation}
dM=T\, dS_+ + V dP + \sum_i \mu_i dq_i,
\end{equation}
where $\mu$ is the chemical potential associated with the charges $q_i = q_m, q_e$.
Applying the first law using the given expressions for mass and temperature yields an expression for the associated entropy. { To calculate entropy we follow a simple method used in Ref.s \cite{newref01, newref02, newref03}. The entropy of the black hole is found to be }
\begin{equation}
S_+=\pi\,r_+^2+4\alpha\, log\left[r_+\right].
\end{equation}
The quantity heat capacity function may be considered in order to analyze the local stability of dyonic $AdS$ black holes with quasitopological electromagnetism. This function is  given by \cite{heat1}
\begin{equation}
C_+=\frac{\partial M}{\partial T}=\bigg(\frac{\partial M}{\partial r_+}\bigg)\bigg(\frac{\partial r_+}{\partial T}\bigg),
\end{equation}
or explicitly in terms of the parameter space, the function is expressed as
\begin{equation}
C_+=-\frac{C_1}{C_2}\hspace{1cm},
\end{equation}
with
\begin{eqnarray}
C_1&=&2 \pi \mathcal{W} \bigg(32 \beta\, q_m^2 \left(l^2 \left(4 \alpha +q_m^2-4 r_+^2\right)-12 r_+^4\bigg)+\ell^2 r_+^4
   \bigg(4 \alpha +q_e^2+q_m^2-4 r_+^2\right)-12 r_+^8\bigg)\nonumber\\
   C_2&=&r_+^8\bigg(12(r_+^6+6r_+^4\alpha)+\ell^2\mathcal{X}\bigg)+1024\beta^2\,q_m^4\bigg(12(r_+^6+6r_+^4\alpha)+\ell^2\mathcal{Y}\bigg)+32\beta\,q_m^2r_+^4\bigg(24(r_+^6\nonumber\\
   &+&6r_+^4\alpha)+\ell^2\mathcal{Z}\bigg)\nonumber.
\end{eqnarray}
where
\begin{eqnarray}
\mathcal{W}&=&\left(2 \alpha +r_+^2\right)^2 \left(32 \beta\, q_m^2+r_+^4\right)\\
\mathcal{X}&=&8 \alpha ^2+2 \alpha  \left(q_e^2+q_m^2+10 r_+^2\right)+3 r_+^2 \left(q_e^2+q_m^2\right)-4 r_+^4\\
\mathcal{Y}&=&8 \alpha ^2+q_m^2 \left(2 \alpha +3 r_+^2\right)-4 r_+^4+20 \alpha  r_+^2\\
\mathcal{Z}&=&16 \alpha ^2-q_e^2 \left(6 \alpha +r_+^2\right)+4 \alpha  q_m^2+6 q_m^2 r_+^2-8 r_+^4+40 \alpha  r_+^2.
\end{eqnarray}
Within certain limits, the associated heat capacity reduces to standard Schwarzschild's case, as noted earlier in the comments on subsequent thermodynamic quantities such as mass and temperature. To check this stability graphically, we depicted the variation of the heat capacity as a function of the horizon radius in Fig. \ref{HC2}. Roughly speaking, the main essence of the ``heat capacity" expression emerges from its sign, i.e., either a positive or negative sign. 

In this regard, Fig. \ref{HC2} depicts the heat capacity behavior with respect to the horizon radius $r_+$. In particular, this variation is driven only by the change of the parameters $q_m$, $q_e$, and $\alpha$. Clearly, the behavior of the heat capacity involves three subregions separated by the existence of two critical radii only for a particular choice of the parameters $q_m$, $q_e$, and $\alpha$. By the way, the same behavior is depicted when the given parameters change. As far as small horizon radii are concerned, the heat capacity is positive, producing a stable small black hole. Next, the heat capacity will be negative after crossing the first subregion, providing an unstable intermediate black hole. The heat capacity quickly changes its sign to be positive once again, which proves that the large black hole is locally stable. It is worth mentioning that, for a certain fixed value of the magnetic charge, the singular points involving the change of sign of the heat capacity are completely hidden. As a result, when the magnetic charge is varied, the critical radii $r_{c1}$ (radius corresponding to first critical point) and $r_{c2}$ (radius corresponding to second critical point) exhibit opposite behavior. At a small horizon radius $r_+$, $r_{c1}$ is fully proportional to the variation ($q_m$, $q_e$, $\alpha$). At a large horizon radius, however, as ($q_m$, $q_e$, $\alpha$) increases, $r_{c2}$ decreases. { As a contribution to the gravity sector, the dyonic charges, which comprise of both magnetic and electric charges, serve as the contributing parameters in understanding the impact of the matter source. Notably, the phenomenon of a second-order phase transition occurs across all scenarios discussed in relation to heat capacity. Specifically, the effect of matter charges plays a significant role in this regard, as it leads to the occurrence of a second-order phase transition in black hole results. A comprehensive investigation is imperative to uncover other essential characteristics associated with the thermodynamics of such black holes which we leave as a future prospect of this study.}

}

\section{Electromagnetic Quasinormal modes}\label{sec3}
\begin{figure}

\begin{minipage}{.539\linewidth}
\centering
\includegraphics[scale=.4]{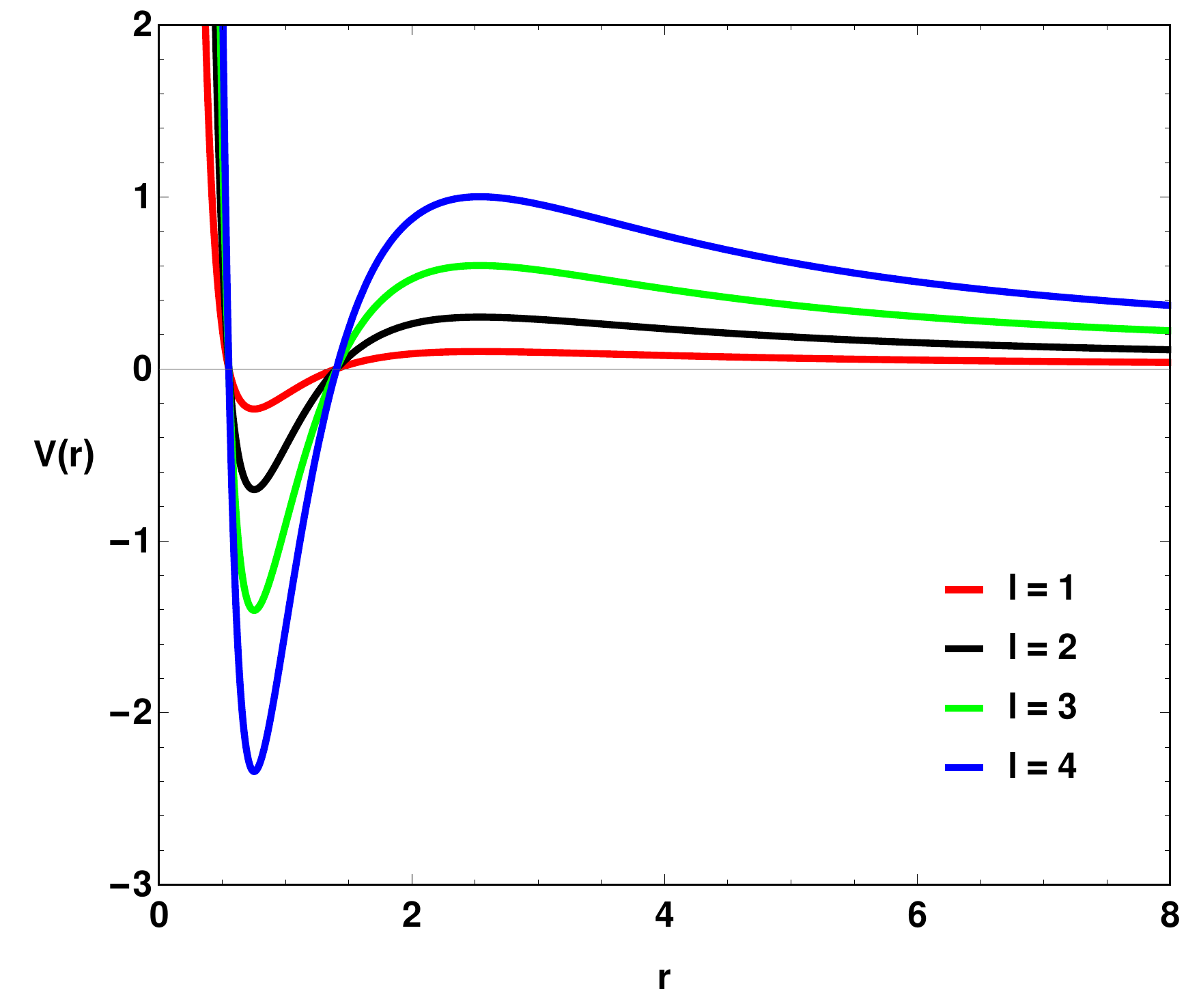}
\end{minipage}%
\begin{minipage}{.5\linewidth}
\centering
\includegraphics[scale=.4]{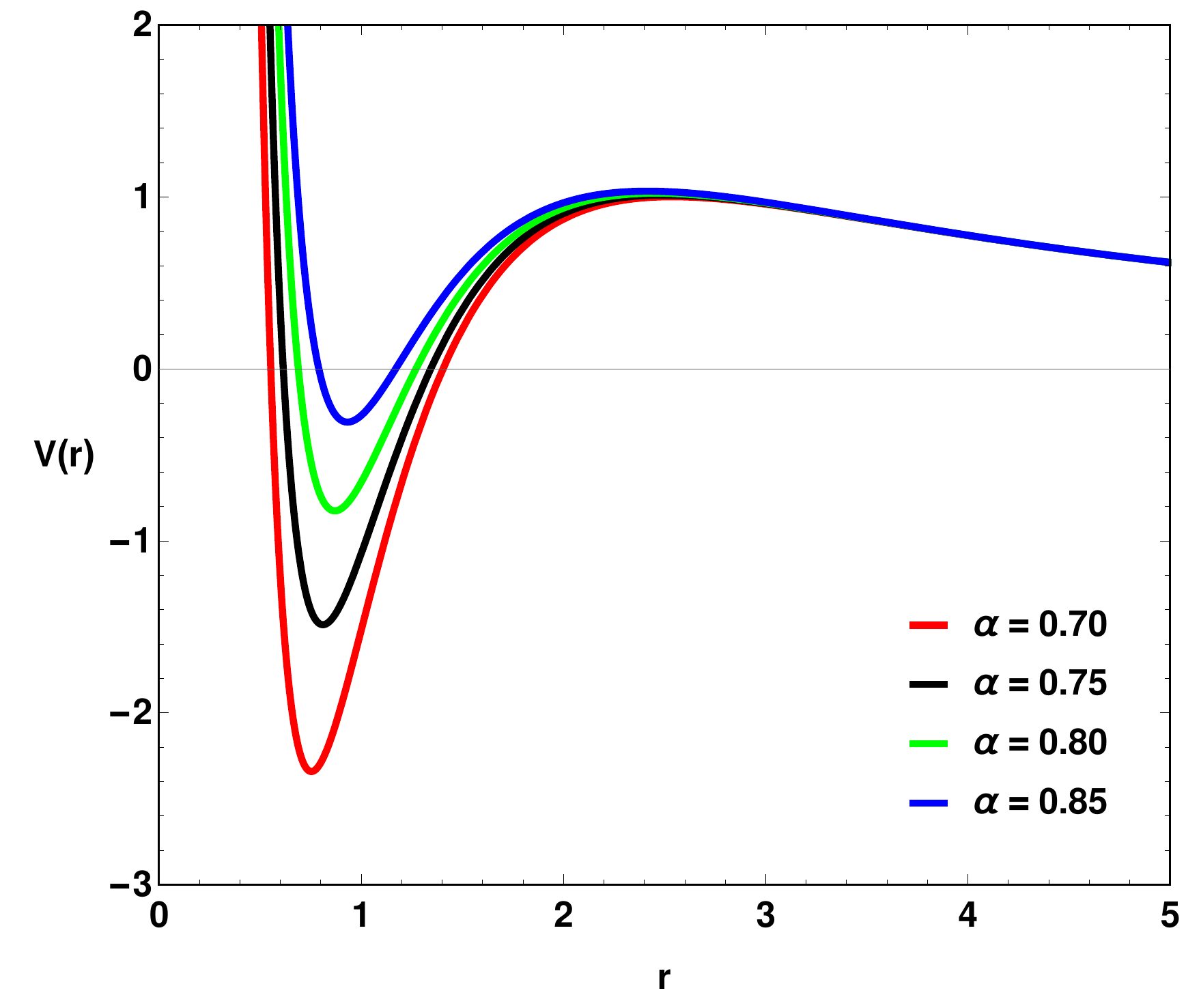}
\end{minipage}\par\medskip

\caption{\label{fig:V01} Behaviour of the potential associated with electromagnetic perturbation. Left panel (a): for different values of multipole moment $l$ with $q_m= 0.6$, $q_e=0.3$, $\alpha = 0.7 $, $\beta = 0.3$, $\Lambda = -0.02$ and $M = 1$  and right panel (b): for different values of $\alpha$ with $q_m= 0.6$, $q_e=0.3$, $\beta = 0.3$, $\Lambda = -0.02$, $M = 1$ and $l=4$. }
\end{figure}

\begin{figure}

\begin{minipage}{.539\linewidth}
\centering
\includegraphics[scale=.4]{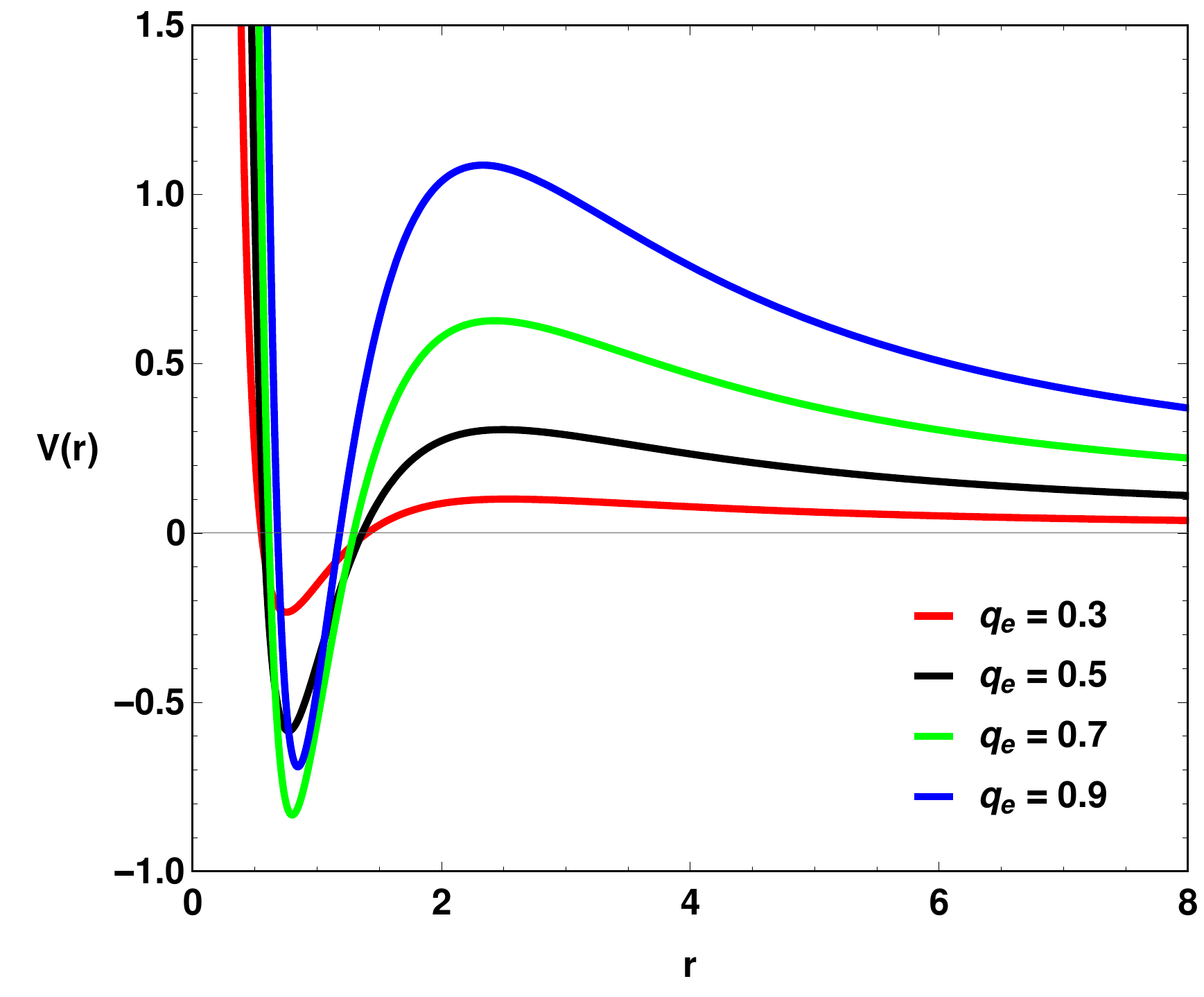}
\end{minipage}%
\begin{minipage}{.5\linewidth}
\centering
\includegraphics[scale=.4]{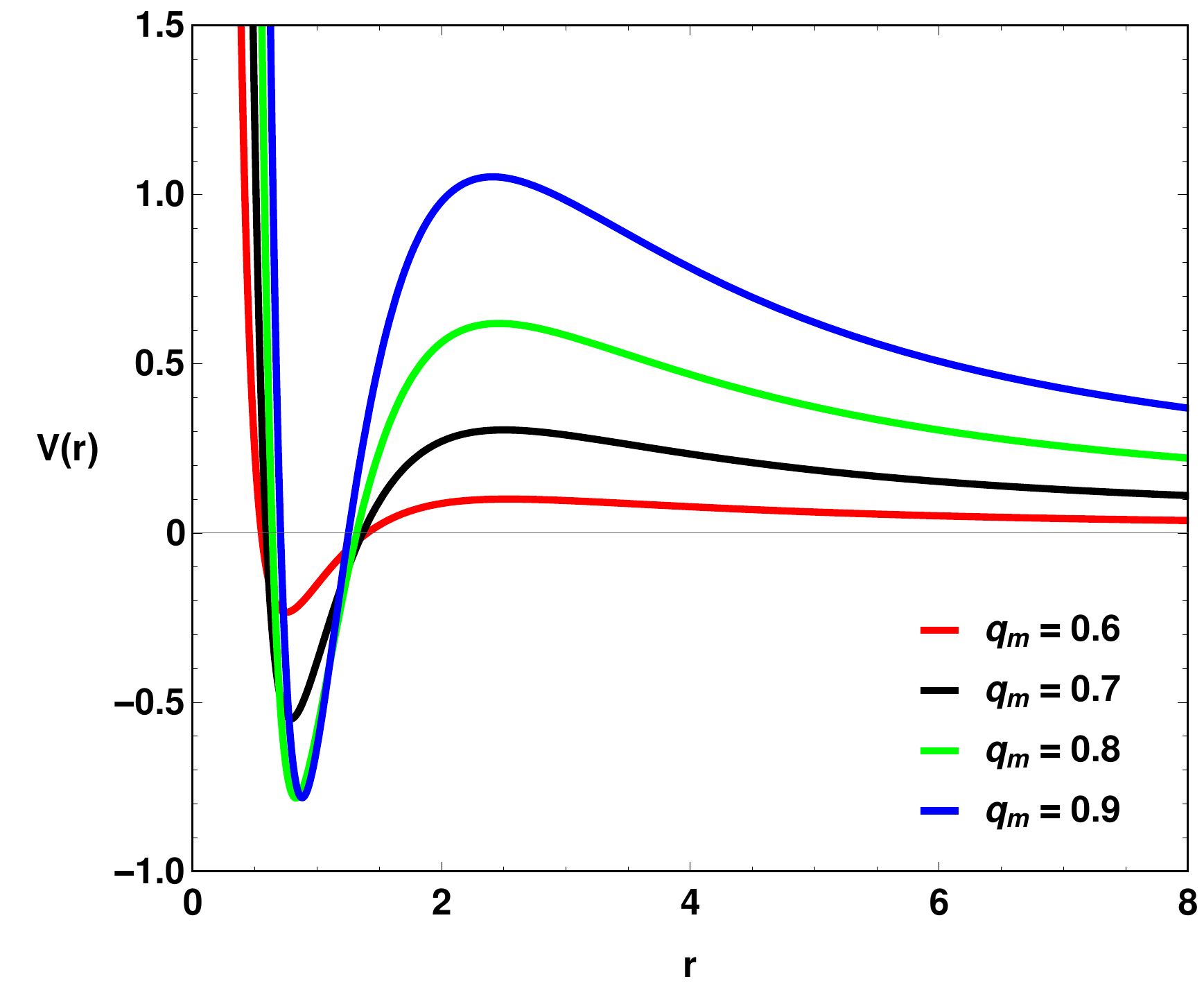}
\end{minipage}\par\medskip

\caption{\label{fig:V02} Behaviour of the potential associated with electromagnetic perturbation. Left panel (a): for different values of $q_e$ with $q_m = 0.6$, $l = 4$, $\alpha = 0.7 $, $\beta = 0.3$, $\Lambda = -0.02$ and $M = 1$  and right panel (b): for different values of $q_m$ with $\alpha= 0.7$, $q_e=0.3$, $\beta = 0.3$, $\Lambda = -0.02$, $M = 1$ and $l=4$.}
\end{figure}

\begin{figure}

\centering
\includegraphics[scale=.4]{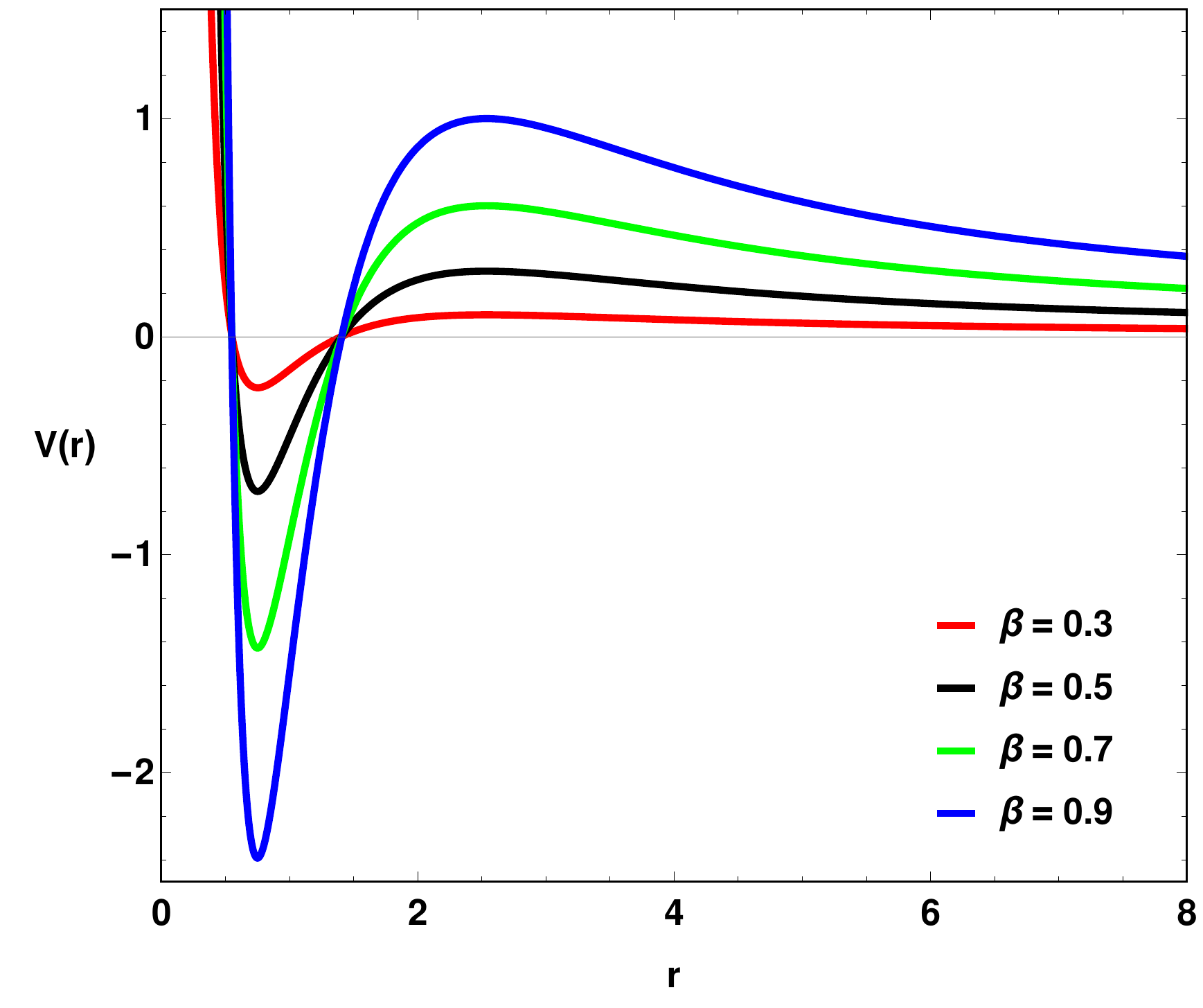}

\caption{\label{fig:V03} Behaviour of the potential associated with electromagnetic perturbation for different values of $\beta$ with $q_m= 0.6$, $q_e=0.3$, $\alpha = 0.7 $, $l = 4$, $\Lambda = -0.02$ and $M = 1$.}
\end{figure}

This section takes care to study the quasinormal modes' behaviour regarding our black hole solution. We shall consider electromagnetic perturbation in the black hole background to study the quasinormal modes.
In case of electromagnetic perturbation, we use the tetrad 
formalism \cite{chandrasekhar, lopez2020, 22a}. Here the basis $e^\mu_{a}$ is defined 
associated with the metric $g_{\mu\nu}$ which should satisfy,
\begin{align}
e^{(a)}_\mu e^\mu_{(b)} &= \delta^{(a)}_{(b)} \notag \\
e^{(a)}_\mu e^\nu_{(a)} &= \delta^{\nu}_{\mu} \notag \\
e^{(a)}_\mu &= g_{\mu\nu} \eta^{(a)(b)} e^\nu_{(b)}\notag \\
g_{\mu\nu} &= \eta_{(a)(b)}e^{(a)}_\mu e^{(b)}_\nu = e_{(a)\mu} e^{(a)}_\nu.
\end{align}
Now, the tensor fields can be expressed as
\begin{align*}
S_\mu &= e^{(a)}_\mu S_{(a)}, \\ 
S_{(a)} &= e^\mu_{(a)} S_\mu, \\
P_{\mu\nu} &=  e^{(a)}_\mu e^{(b)}_\nu P_{(a)(b)}, \\
P_{(a)(b)} &= e^\mu_{(a)} e^\nu_{(b)} P_{\mu\nu}.
\end{align*}
In the tetrad formalism the covariant derivative in the 
actual coordinate system is replaced with the intrinsic derivative in the 
tetrad frame as shown below \cite{chandrasekhar, lopez2020}:
\begin{align}
K_{(a)(b)|(c)} &\equiv e^{\lambda}_{(c)} K_{\mu\nu;\lambda} e^\mu_{(a)} e^\nu_{(b)} \notag \\
&= K_{(a)(b),(c)} - \eta^{(m)(n)} (\gamma _{(n)(a)(c)} K_{(m)(b)} + \gamma _{(n)(b)(c)} K_{(a)(m)}),
\end{align}
where the Ricci rotation coefficients are expressed by $\gamma_{(c)(a)(b)} \equiv e^\mu_{(b)} e_{(a)\nu;\mu} e^\nu_{(c)}.$ The vertical rule and the comma denote 
the intrinsic and directional derivative respectively in the tetrad basis. 
Now for the electromagnetic perturbation in the tetrad formalism, the Bianchi 
identity of the field strength $F_{[(a)(b)(c)]} = 0$ gives
\begin{align}
\left( r \sqrt{f(r)}\, F_{(t)(\phi)}\right)_{,r} + r \sqrt{f(r)^{-1}}\, F_{(\phi)(r), t} &=0, \label{em1} \\
\left( r \sqrt{f(r)}\, F_{(t)(\phi)}\sin\theta\right)_{,\theta} + r^2 \sin\theta\, F_{(\phi)(r), t} &=0. \label{em2}
\end{align}
The conservation equation is
\begin{equation}
\eta^{(b)(c)}\! \left( F_{(a)(b)} \right)_{|(c)} =0.
\end{equation}
The above equation can be further written as
\begin{equation} \label{em3}
\left( r \sqrt{f(r)}\, F_{(\phi)(r)}\right)_{,r} +  \, F_{(\phi)(\theta),\theta} + r \sqrt{f(r)^{-1}}\, F_{(t)(\phi), t} = 0.
\end{equation}
Differentiating equation \eqref{em3} w.r.t.\ $t$ and using equations \eqref{em1} and \eqref{em2}, we get,
\begin{equation}\label{em4}
\left[ f(r) \left( r \sqrt{f(r)}\, \mathcal{F} \right)_{,r} \right]_{,r} + \dfrac{ \sqrt{f(r)}}{r} \left( \dfrac{\mathcal{F}_{,\theta}}{\sin\theta} \right)_{,\theta}\!\! \sin\theta - r \sqrt{f(r)^{-1}}\, \mathcal{F}_{,tt} = 0,
\end{equation}
where we have considered $\mathcal{F} = F_{(t)(\phi)} \sin\theta.$ Using the 
Fourier decomposition $(\partial_t \rightarrow -\, i \omega)$ and field 
decomposition $\mathcal{F}(r,\theta) = \mathcal{F}(r) Y_{,\theta}/\sin\theta,$ where $Y(\theta)$ is the Gegenbauer function and it satisfies the following 
relation,
\begin{equation}
\sin\theta\, \dfrac{d}{d\theta} \left( \dfrac{1}{\sin\theta} \dfrac{d}{d\theta} \dfrac{Y_{,\theta}}{\sin\theta} \right) = -l(l+1) \dfrac{Y_{,\theta}}{\sin\theta},
\end{equation}
we can write equation \eqref{em4} in the following form:
\begin{equation}\label{em5}
\left[ f(r) \left( r \sqrt{f(r)}\, \mathcal{F} \right)_{,r} \right]_{,r} + \omega^2 r \sqrt{f(r)^{-1}}\, \mathcal{F} -  \sqrt{f(r)} r^{-1} l(l+1)\, \mathcal{F} = 0.
\end{equation}
Now, finally using the tortoise coordinate defined by 
\begin{equation}
\dfrac{dr_*}{dr} = \dfrac{1}{f(r)},
\end{equation} 
and using $\psi_e \equiv r \sqrt{f(r)}\, \mathcal{F}$, equation \eqref{em5} can be expressed as the Schr\"odinger like form given by the following equation
\begin{equation} \label{Scheqn}
\partial^2_{r_*} \psi_e + \omega^2 \psi_e = V(r) \psi_e,
\end{equation}
where the potential is,
\begin{equation}\label{Ve}
V(r) = f(r)\, \dfrac{l(l+1)}{r^2},
\end{equation}
where $l$ denotes  the multipole number. To inspect the scenarios physically, we should treat the boundary conditions on the wave function at the event horizon as well as at spatial infinity. Indeed, neither at the event horizon nor at spatial infinity does a signal come out to excite the black hole even more. Therefore, a suitable selection of boundary conditions ensures that waves are outgoing at both boundaries. This can be expressed as follows:
\begin{equation}
\Phi=\left\{
\begin{array}{lr}
e^{i \omega r_*} \  \ \text{if} \   \ r_*\rightarrow-\infty\\
e^{- i \omega r_*} \  \ \text{if} \   \ r_*\rightarrow+\infty\end{array}
\right.
\label{Eq31}
\end{equation}
%\begin{center}
%\textbf{illustration and discussion of the potential behavior}
%\end{center}
Now, before going to the analysis of quasinormal modes, we study the behaviour of the electromagnetic perturbation in the black hole hole spacetime. In Fig. \ref{fig:V01}, on the left panel, we have plotted the potential $V$ w.r.t. $r$ for different values of the multipole no. $l$ with the model parameters $q_m= 0.6$, $q_e=0.3$, $\alpha = 0.7 $, $\beta = 0.3$, $\Lambda = -0.02$ and $M = 1$. The potential shows suitable behaviour and the peak of the curve increases with an increase in the value of multipole no. $l$. However, for small values of $r$, we observe a negative peak of the potential which gets smaller with an increase in the value of $l$. From the behaviour of potential, we can see that the peak is well distinguished even for $l=1$ and WKB method which is highly connected with the peak of the potential should be applicable to this case from $l=1$. On the right panel of Fig. \ref{fig:V01}, we have plotted the potential $V(r)$ w.r.t. $r$ for different values of model parameter $\alpha$. Here we see that the impact of $\alpha$ on the peak of the potential is relatively small but the negative peak is comparatively more sensitive to the parameter $\alpha$. In Fig. \ref{fig:V02}, we have shown the variation of the potential for different values of $q_e$ (on the left panel) and $q_m$ (on the right panel). In both cases, we observe a similar behaviour of the potential. Hence, the variation of quasinormal modes may be similar to each other which will be clarified in the next part of the study. Finally, we have plotted the potential $V(r)$ vs. $r$ for different values of $\beta$ in Fig. \ref{fig:V03}. Here also, we see that with an increase in the value of the parameter $\beta$, the peak of the potential increases significantly.

\subsection{WKB Approximation}
The solution of Eq. \ref{Scheqn} is such a major problem in physics that several numerical and semi-analytical methods are used to solve it. In particular, the WKB method is the employed tool for describing the spectrum provided by the quasinormal modes study. Moreover, the application of this method offers us the use of sixth-order corrections for the WKB approximation in the following simple way
\begin{equation}
\frac{i(\omega^2-V_0)}{\sqrt{-2V_0^{''}}}-\sum_{j=2}^{6}\Lambda_j=n+\frac{1}{2},
\label{Eq32}
\end{equation}
where $V_0$ is the maximized potential, $V_0^{''}$ represents a second derivative with respect to the tortoise coordinate,  and $\Lambda_j$ are the correction terms of the method.

\begin{table}
\caption{ Quasinormal modes for $l=4$, $q_m= 0.6$, $\beta=0.3$, $\alpha = 0.4$ $\Lambda = -0.02$ and $M = 1$ for different values of $q_e$ using $3$rd order and $5$th order WKB approximation method.}
\begin{tabular}{|c|c|c|c|c|}
\hline 
$q_e$ & 3rd order WKB & 5th order WKB & $\Delta_3$ & $\Delta_5$ \\ 
\hline 
$0.01$ & $0.9640884943823855-i 0.09534849108657481$ & $0.964254686280287 -i 0.09537015845275655$ & $0.000640682$ & $0.0000226069$ \\ 
\hline 
$0.03$ & $0.9641223512533119 -i 0.09534763055225685$ & $0.9642884300487975 -i 0.09536828277600931$ & $0.000640629$ & $0.000365657$ \\ 
\hline 
$0.05$ & $0.9641900844172705 -i 0.09534590724753911$ & $0.9643560361839089 -i 0.09536552840694257$ & $0.000640524$ & $0.0000331457$ \\ 
\hline 
$0.07$  & $0.9644575107684801 -i 0.09536167761755494$ & $0.9642917327933695 -i 0.09534331710035836$ & $0.000640365$ & $0.000101375$ \\ 
\hline 
$0.09$ & $0.9644273548378068 -i 0.09533985383563146$ & $0.964593224470319 -i 0.09535963467292152$ & $0.000640154$ & $0.000018$ \\ 
\hline 
\end{tabular} \label{table01}
\end{table}

\begin{figure}

\begin{minipage}{.539\linewidth}
\centering
\includegraphics[scale=.4]{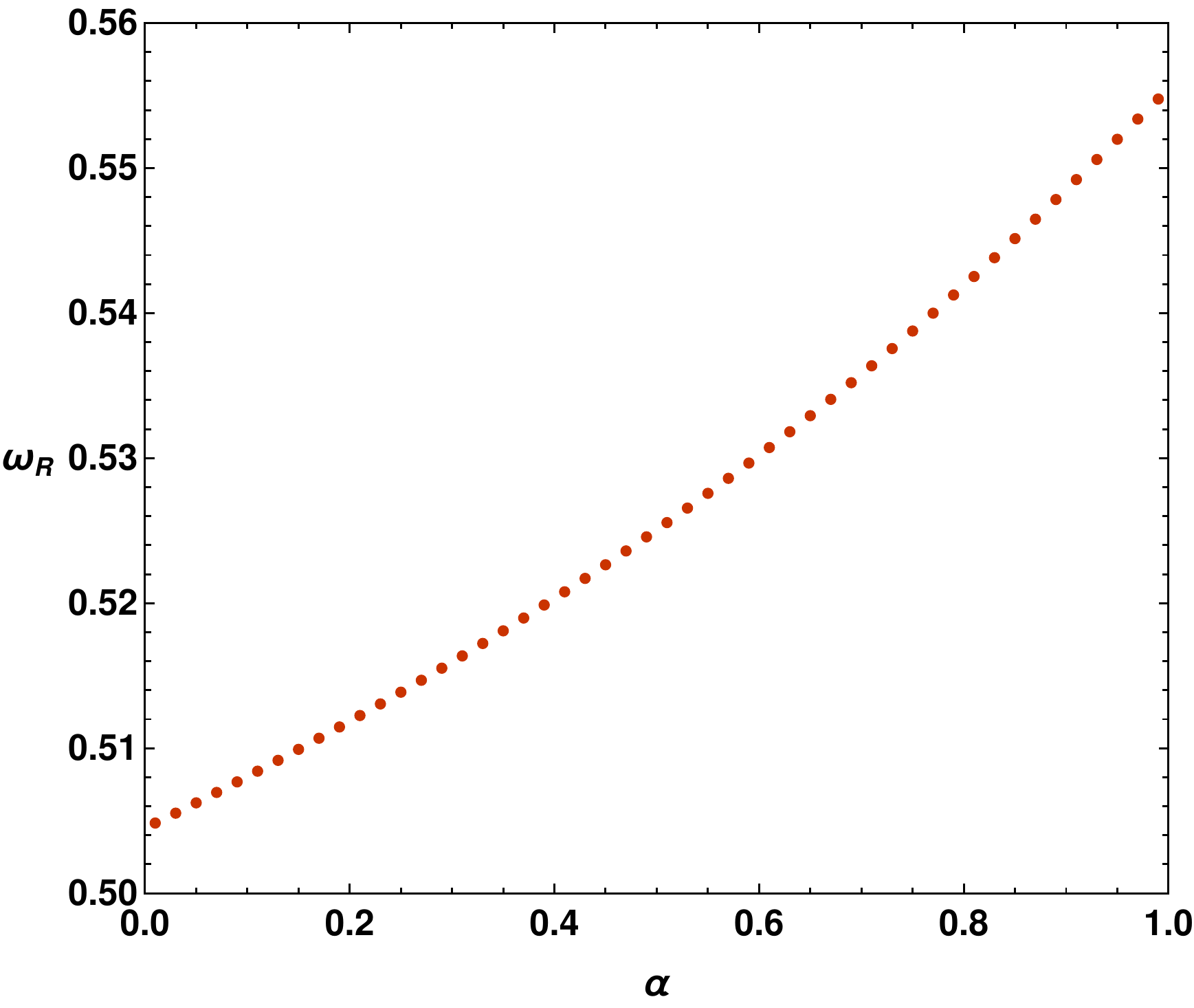}
\end{minipage}%
\begin{minipage}{.5\linewidth}
\centering
\includegraphics[scale=.4]{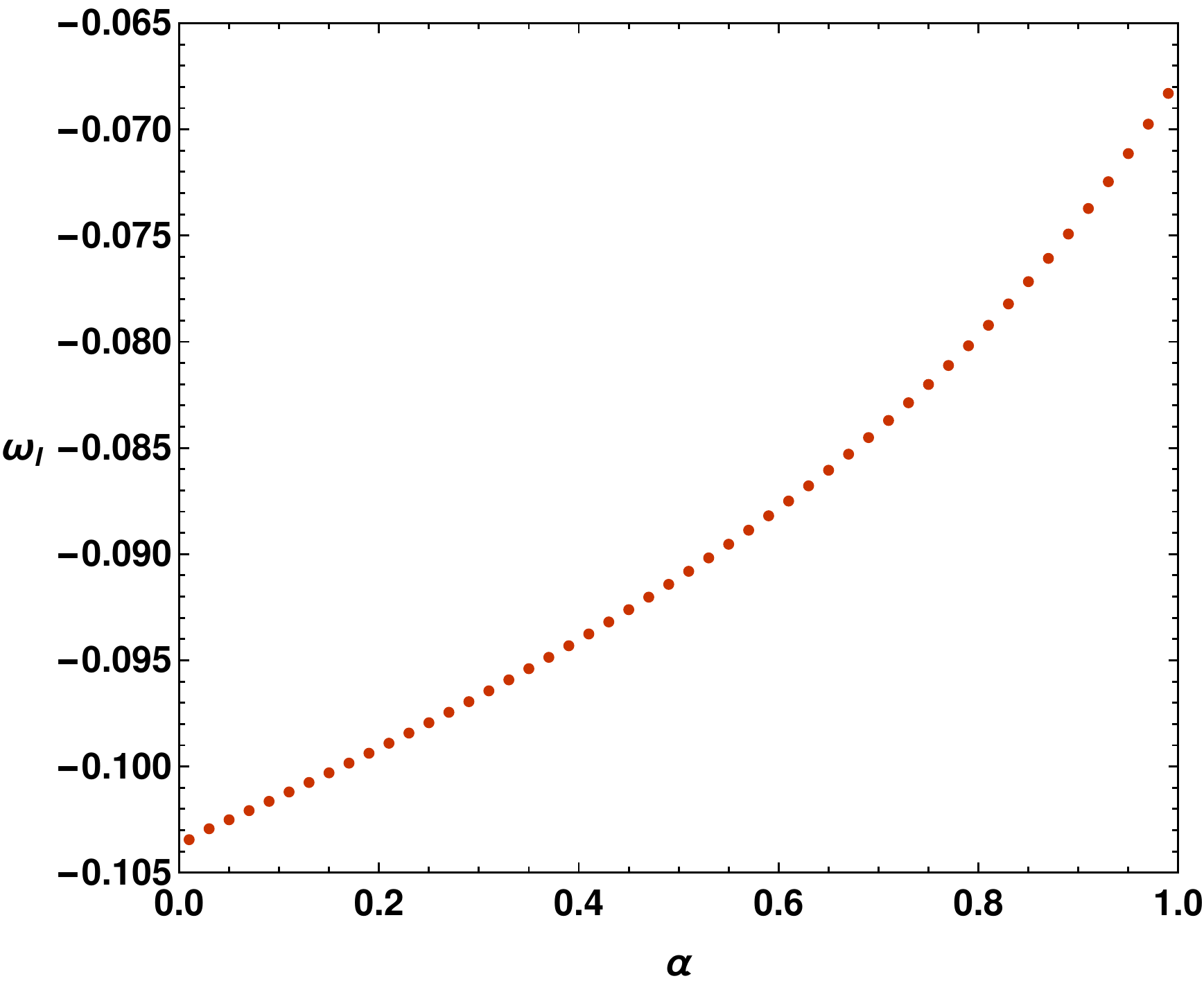}
\end{minipage}\par\medskip

\caption{\label{fig:QNM01} Quasinormal modes for $l=2$, $q_m= 0.6$, $q_e=0.3$, $\beta = 0.3$, $\Lambda = -0.02$, $M = 1$. Left panel (a): real quasinormal modes and right panel (b): imaginary quasinormal modes.}
\end{figure}

We have calculated the quasinormal modes using higher order WKB approximation method and listed some results in the Table \ref{table01} for different values of $q_e$. In the Table \ref{table01}, $$\Delta_3 = \dfrac{|\omega_4 - \omega_2|}{2},$$ and $$\Delta_5 = \dfrac{|\omega_6 - \omega_4|}{2}.$$ $\omega_i$ represents quasinormal modes calculated using $i$th order WKB approximtion method. The term $\Delta_i$ represents error associated with $i$th order WKB approximation method \cite{22a}. From Table \ref{table01}, one can see that the errors associated with 3rd order WKB approximation are comparatively higher. Hence, we have used 5th order WKB method for the graphical analysis part. In Fig. \ref{fig:QNM01}, we have plotted the real quasinormal modes and the imaginary quasinormal modes vs. $\alpha$ on the left and right panel respectively. One can see that with an increase in the parameter $\alpha$ from $0$ to $1$, the real quasinormal frequencies increase gradually in a non-linear pattern. While on the other hand, increase in the parameter $\alpha$ decreases the decay rate of gravitational waves. Hence, an increase in this parameter $\alpha$ increases the energy (frequency) of gravitational waves and the decay rate also decreases ensuring longer life time of the propagating gravitational waves.

\begin{figure}

\begin{minipage}{.539\linewidth}
\centering
\includegraphics[scale=.4]{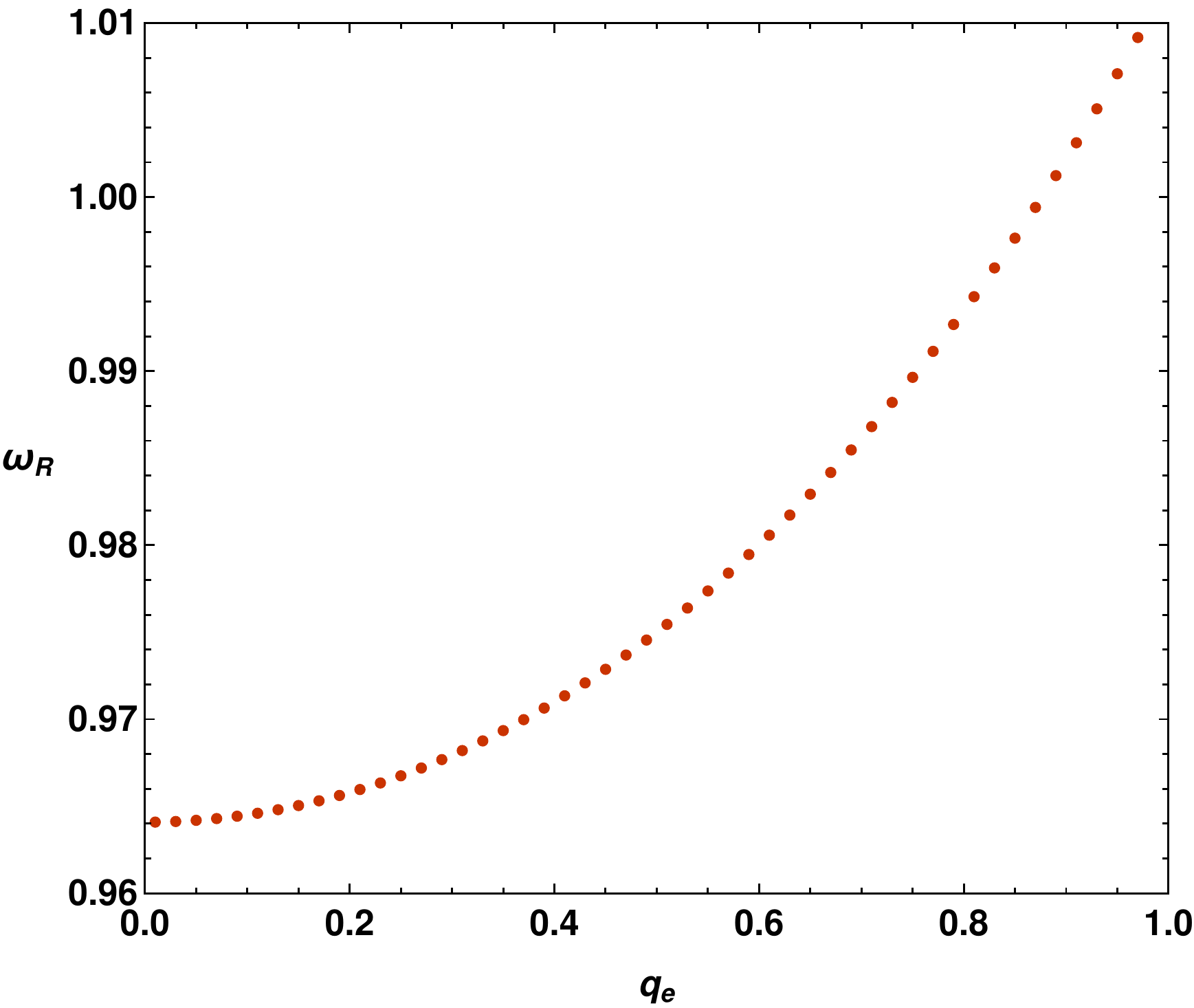}
\end{minipage}%
\begin{minipage}{.5\linewidth}
\centering
\includegraphics[scale=.4]{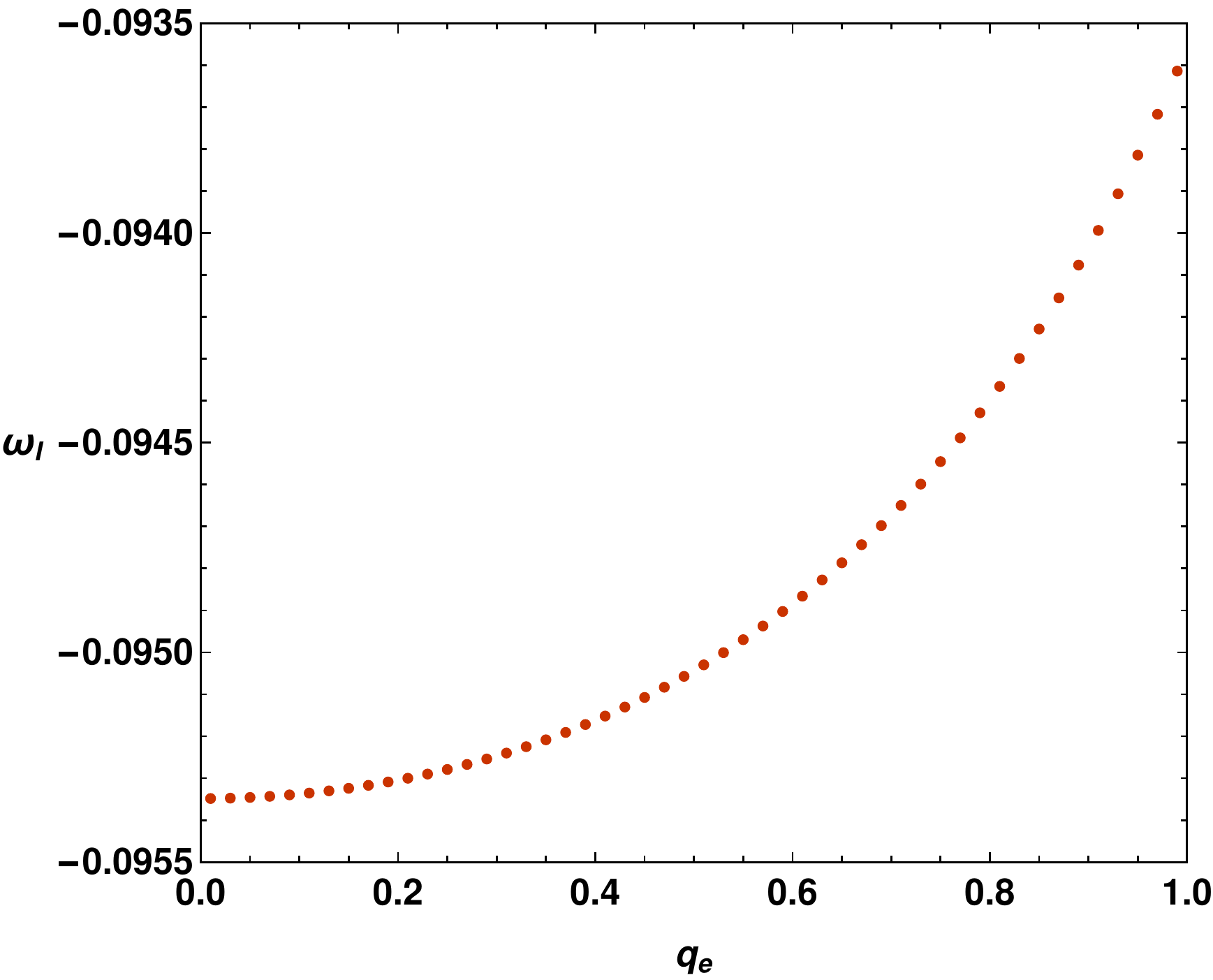}
\end{minipage}\par\medskip

\caption{\label{fig:QNM02} Quasinormal modes for $l=4$, $q_m= 0.6$, $\beta=0.3$, $\alpha = 0.4$ $\Lambda = -0.02$, $M = 1$. Left panel (a): real quasinormal modes and right panel (b): imaginary quasinormal modes.}
\end{figure}

In Fig. \ref{fig:QNM02}, we have shown the variation of real and imaginary quasinormal modes with respect to the parameter $q_e$ i.e. the electric charge of the black hole. The left panel shows the quasinormal frequencies while the right panel shows the decay rate of the waves. One can see that the impacts of the parameter $q_e$ is not identical to that of the previous case for variation of quasinormal modes with respect to $\alpha$. Here, the variation of quasinormal frequencies are more non-linear. For the smaller values of $q_e$ near $0$, the variation is very small but it increases in an exponential form when $q_e$ approaches $1$. The decay rate also decreases exponentially when $q_e$ increases to $1$. 

\begin{figure}

\begin{minipage}{.539\linewidth}
\centering
\includegraphics[scale=.4]{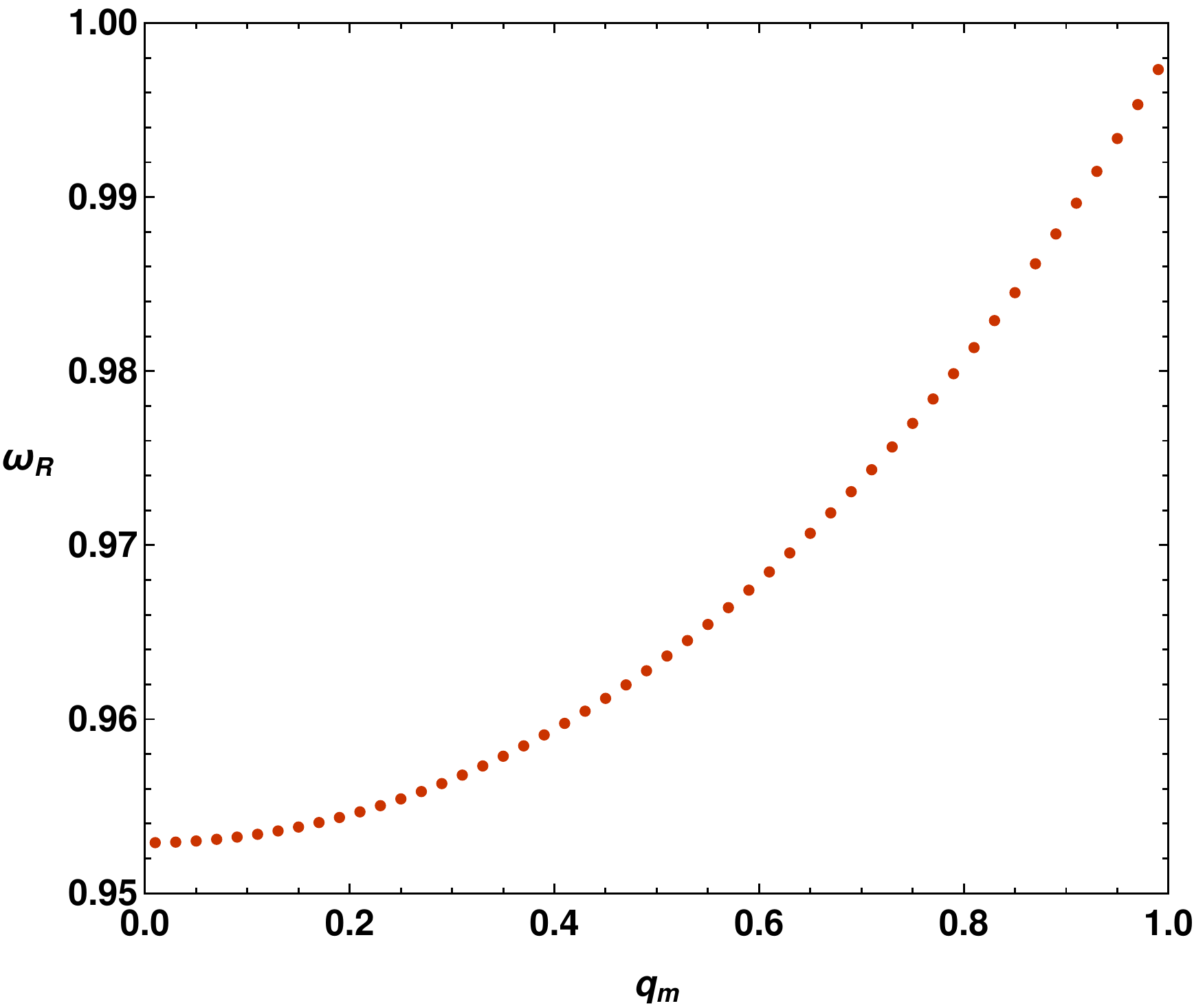}
\end{minipage}%
\begin{minipage}{.5\linewidth}
\centering
\includegraphics[scale=.4]{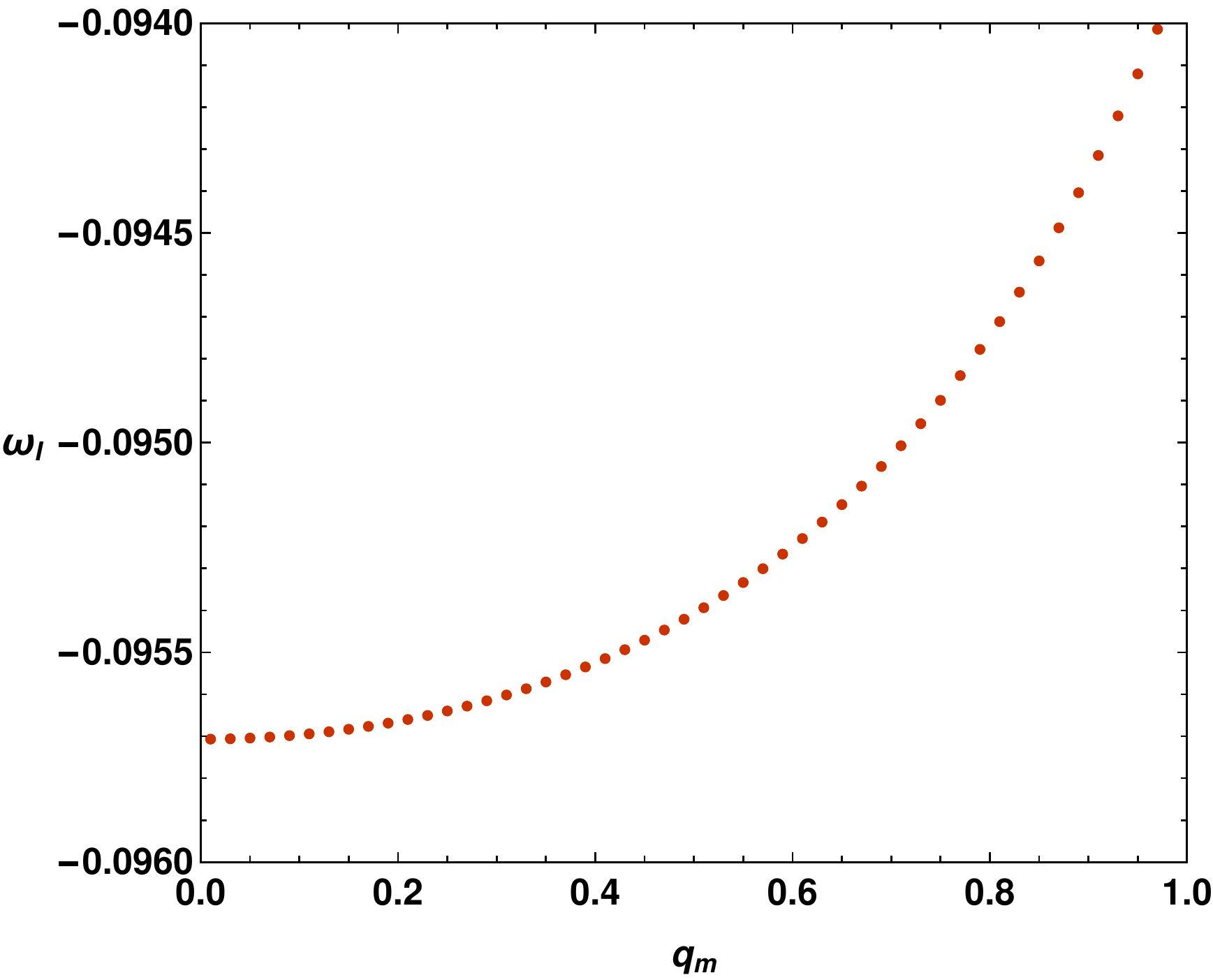}
\end{minipage}\par\medskip

\caption{\label{fig:QNM03} Quasinormal modes for $l=4$, $\beta= 0.3$, $q_e=0.3$, $\alpha = 0.4$, $\Lambda = -0.02$, $M = 1$. Left panel (a): real quasinormal modes and right panel (b): imaginary quasinormal modes.}
\end{figure}

\begin{figure}

\begin{minipage}{.539\linewidth}
\centering
\includegraphics[scale=.4]{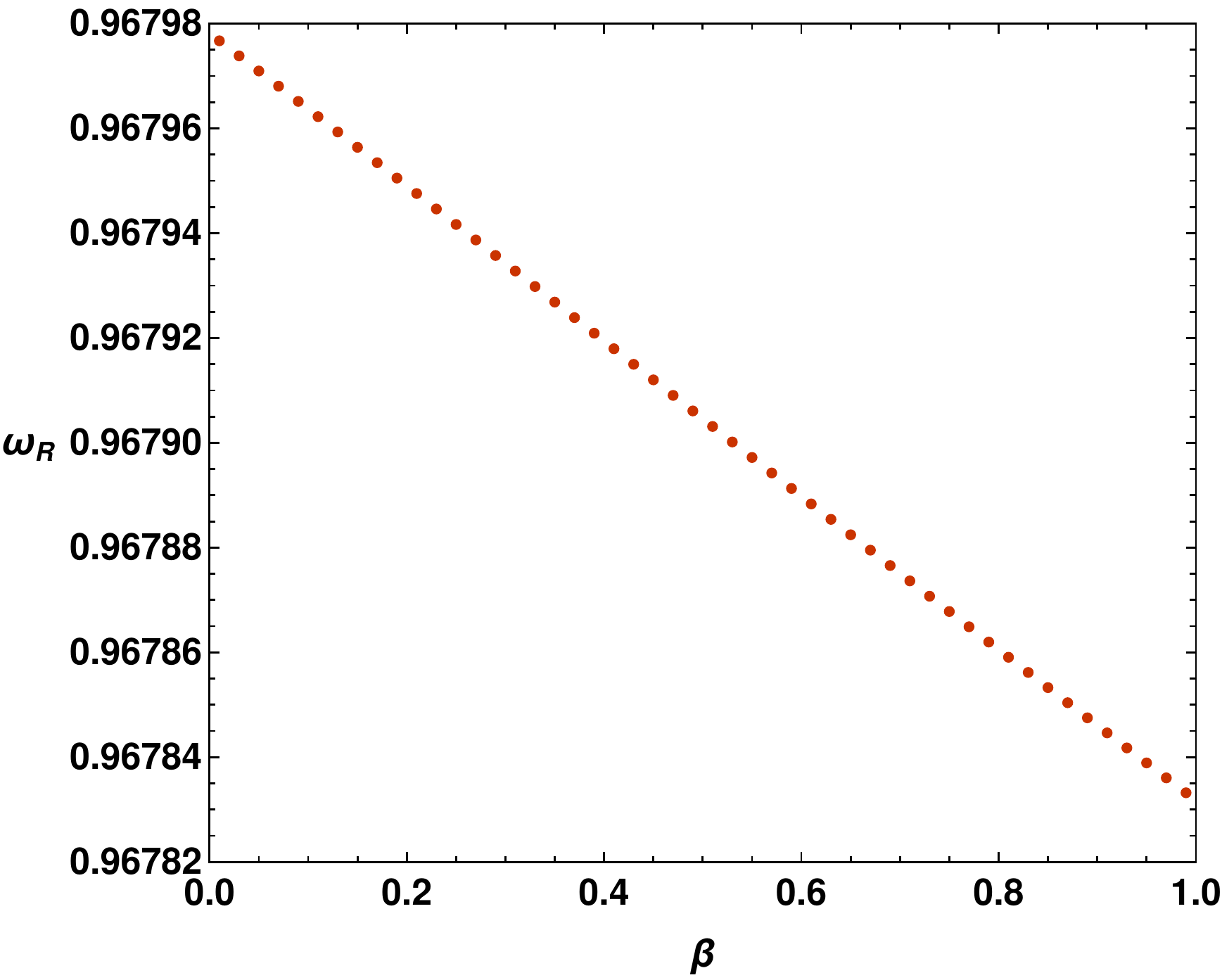}
\end{minipage}%
\begin{minipage}{.5\linewidth}
\centering
\includegraphics[scale=.4]{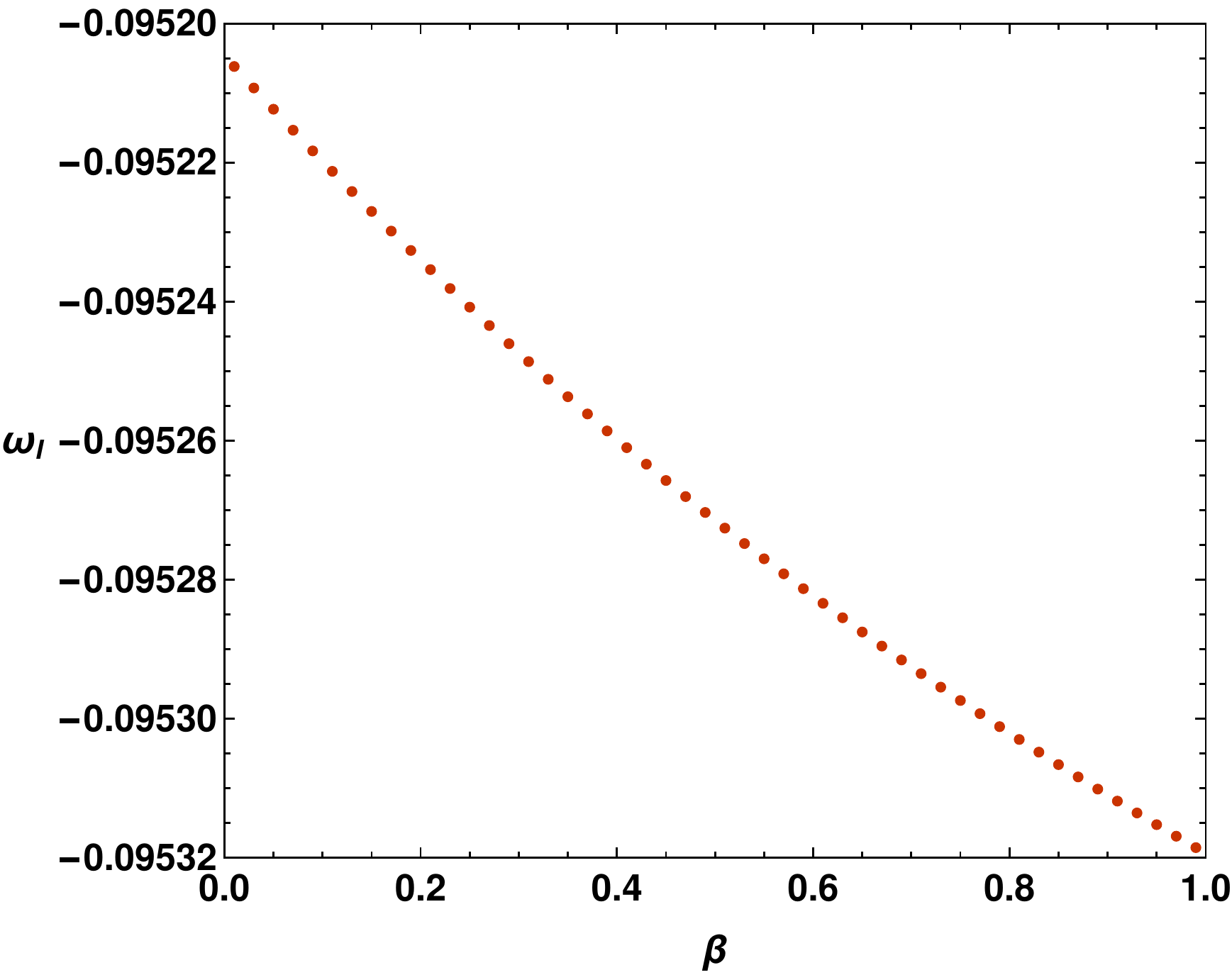}
\end{minipage}\par\medskip

\caption{\label{fig:QNM04} Quasinormal modes for $l=4$, $q_m= 0.6$, $q_e=0.3$, $\alpha = 0.4$, $\Lambda = -0.02$, $M = 1$. Left panel (a): real quasinormal modes and right panel (b): imaginary quasinormal modes.}
\end{figure}

In Fig. \ref{fig:QNM03}, we have shown the variation of quasinormal modes w.r.t. $q_m$. Here, one can see that the impact of parameter $q_m$ on quasinormal modes is identical to the impact of $q_e$. Finally, we have studied the quasinormal modes w.r.t. model parameter $\beta$ in Fig. \ref{fig:QNM04}. On the left panel of Fig. \ref{fig:QNM04}, we have shown the variation of real quasinormal frequencies w.r.t. $\beta$. One can see that with an increase in $\beta$, the oscillation frequencies of gravitational waves decrease following almost a linear pattern. However, the decay rate of gravitational waves increases with an increase in the parameter $\beta$ (see the right panel of Fig. \ref{fig:QNM04}).

%\begin{center}
%\textbf{Part dedicated to the discussion of QNMs by WKB method}
%\end{center}
\section{Conclusion}\label{sec4}
In this paper, we derived a black hole solution in the framework of EGB gravity coupled to a matter source represented by a quasi-topological electromagnetism term providing both magnetic and electric charge. This solution recovered some limits to the usual black hole solution, especially in the context of general relativity when the GB coupling constant was close to zero.

A more in-depth investigation of the electromagnetic quasinormal modes process by a perturbation over the black hole spacetime was conducted. Revealing the spectrum of the perturbation analytically is too complicated, so numerical methods are exploited. The WKB approximation was used here to extract the complex part as well as the real part of the electromagnetic perturbation. 
The study shows that both the charges i.e. electric and magnetic charges have similar impacts on the quasinormal modes of the black hole. Hence it might be difficult to comment on the type of charges by simply observing the quasinormal modes experimentally in the near future. Another interesting result is observed for the parameter $\beta$. An increase in $\beta$ decreases the real quasinormal modes and increases the decay rate of gravitational waves.
This work raises open questions and suggests future research, such as on shadow behaviors or deflection light behaviors. 

{ Recently in Ref.s \cite{new06, new07, new08}, relation between shadow radius and black hole quasinormal modes have been investigated. For this scenario also, we expect relations between black hole shadow radius and quasinormal modes. Basically, the real quasinormal frequency will depend on the multipole moment $l$ and inverse of the shadow radius of the black hole \cite{new08}. On the other hand, imaginary part depends on the overtone number and inverse of shadow radius \cite{new08}. A detailed investigation dealing with the shadow behaviours with the model parameters will shed more light into such a study. We keep this as a future prospect of the study.
}

\section*{Acknowledgments}
The authors are thankful to the anonymous referee for useful suggestions which has improved the manuscript. DJG is thankful to Prof. U. D. Goswami for some useful discussions. DJG acknowledges the contribution of the COST Action CA21136  -- ``Addressing observational tensions in cosmology with systematics and fundamental physics (CosmoVerse)".


\begin{thebibliography}{0} % 100 is a random guess of the total number of
%references
\bibitem{1a}S. W. Hawking and G. F. R. Ellis. \textit{The Large Scale Structure of Space-Time}. Cambridge Monographs on Mathematical Physics. Cambridge University Press, \textbf{2} 2011.
\bibitem{2a}Natario, J., \textit{Relativity and Singularities: A Short Introduction for Mathematicians}, {\tt arXiv:math/0603190 [math.DG]}.
\bibitem{3a} R. Penrose, \textit{Gravitational collapse and space-time singularities}, Phys. Rev. Lett. \textbf{14} (1965) 57.
\bibitem{4a}K. Akiyama et al. [Event Horizon Telescope], \textit{First M87 Event Horizon Telescope Results. I. The Shadow of the Supermassive Black Hole}, Astrophys. J. Lett. \textbf{875}, L1 (2019), {\tt arXiv:1906.11238 [astro-ph.GA]}.
\bibitem{5a}Event Horizon Telescope collaboration, K. Akiyama et al., \textit{First M87 Event Horizon Telescope Results. VI. The Shadow and Mass of the Central Black Hole}, Astrophys. J. Lett. \textbf{875} (2019) L6, {\tt 	arXiv:1906.11243 [astro-ph.GA]}.
\bibitem{6a}Event Horizon Telescope collaboration, K. Akiyama et al., \textit{First M87 Event Horizon Telescope Results. II. Array and Instrumentation}, Astrophys. J. Lett. 875 (2019) L2, {\tt 	arXiv:1906.11239 [astro-ph.IM]}.
\bibitem{7a}Event Horizon Telescope collaboration, K. Akiyama et al., \textit{First M87 Event Horizon Telescope Results. III. Data Processing and Calibration}, Astrophys. J. Lett. \textbf{875} (2019) L3, {\tt 	arXiv:1906.11240 [astro-ph.GA]}.
\bibitem{8a} Event Horizon Telescope collaboration, K. Akiyama et al., \textit{First M87 Event Horizon Telescope Results. IV. Imaging the Central Supermassive Black Hole}, Astrophys. J. Lett. \textbf{875} (2019) L4,  {\tt 	arXiv:1906.11241 [astro-ph.GA]}.
\bibitem{9a}Event Horizon Telescope collaboration, K. Akiyama et al., \textit{First M87 Event Horizon Telescope Results. V. Physical Origin of the Asymmetric Ring}, Astrophys. J. Lett. \textbf{875} (2019) L5, {\tt 	arXiv:1906.11242 [astro-ph.GA]}.
\bibitem{10a}E. Berti, V. Cardoso, and A. O. Starinets, \textit{Quasinormal modes of black holes and black branes}, Class. Quant. Grav. \textbf{26} (2009) 163001, {\tt 	arXiv:0905.2975 [gr-qc]}.
\bibitem{11a}E. Berti, V. Cardoso, and C. M. Will, \textit{On gravitational-wave spectroscopy of massive black holes with the space interferometer LISA}, Phys. Rev. D \textbf{73} (2006) 064030, {\tt 	arXiv:gr-qc/0512160}.
\bibitem{12a}E. Berti, K. Yagi, H. Yang, and N. Yunes, \textit{Extreme Gravity Tests with Gravitational Waves from Compact Binary Coalescences: (II) Ringdown}, Gen. Rel. Grav. 50 (2018), no. 5 49, {\tt 	arXiv:1801.03587 [gr-qc]}.
\bibitem{13a}Jake Percival and Sam R. Dolan. \textit{Quasinormal modes of massive vector fields on the Kerr spacetime}. Phys. Rev. D, \textbf{102}(10):104055, 2020.
\bibitem{14a} H. T. Cho, A. S. Cornell, Jason Doukas, T. R. Huang, and Wade Naylor. \textit{A New Approach to Black Hole Quasinormal Modes: A Review of the Asymptotic Iteration Method}. Adv. Math. Phys., 2012:281705, 2012.
\bibitem{15a}J. Matyjasek and M. Opala, \textit{Quasinormal modes of black holes: The improved semianalytic approach}, Phys. Rev. D\textbf{96}, 024011 (2017).
\bibitem{16a}Bernard F. Schutz and Clifford M. Will. \textit{Black hole normal modes: a semianalytic approach}. Astrophys. J. Lett., 291:L33–L36, 1985.
\bibitem{17a}D. J. Gogoi and U. D. Goswami, \textit{Quasinormal Modes of Black Holes with Non-Linear-Electrodynamic Sources in Rastall Gravity}, Physics of the Dark Universe \textbf{33}, 100860 (2021).
\bibitem{18a}D. J. Gogoi, R. Karmakar and U. D. Goswami, {\it Quasinormal Modes of Non-Linearly Charged Black Holes surrounded by a Cloud of Strings in Rastall Gravity}, Int. J. Geom. Methods Mod. Phys. {\bf 20}, 2350007 (2023) {\tt arXiv:2111.00854 [gr-qc] (2022)}.
\bibitem{18b}D. J. Gogoi and U. D. Goswami, {\it Tideless Traversable Wormholes surrounded by cloud of strings in f(R) gravity}, JCAP {\bf 02}, 027 (2023).
\bibitem{18c}R. Karmakar, D. J. Gogoi and U. D. Goswami, {\it Quasinormal modes and thermodynamic properties of GUP-corrected Schwarzschild black hole surrounded by quintessence}, Int. J. Mod. Phys. A {\bf 37}, 2250180 (2022).
\bibitem{19a} J. P. M. Graca and I. P. Lobo, \textit{Scalar QNMs for Higher Dimensional Black Holes Surrounded by Quintessence in Rastall Gravity}, Eur. Phys. J. C 78, \textbf{101} (2018) {\tt	arXiv:1711.08714 [gr-qc]}.
\bibitem{20a}M. S. Churilova, \textit{Quasinormal modes of the Dirac field in the consistent 4D Einstein–Gauss–Bonnet gravity}, Phys. Dark Univ. \textbf{31} (2021) 100748.
\bibitem{21a}H. Ma \textit{Quasinormal modes of novel 4D Einstein-Gauss-Bonnet Anti-de Sitter black holes}, EPL \textbf{137} (2022) no.2, 29001.
\bibitem{22a}D. J. Gogoi and U. D. Goswami, \textit{Quasinormal Modes and Hawking Radiation Sparsity of GUP corrected Black Holes in Bumblebee Gravity with Topological Defects}, JCAP \textbf{06} (2022) 029, {\tt arXiv:2203.07594 [gr-qc]}.


\bibitem{heat1}C. H. Nam, {\it Non-Linear Charged DS Black Hole and Its Thermodynamics and Phase Transitions}, Eur. Phys. J. C {\bf 78}, 418 (2018).

\bibitem{chandrasekhar}S. Chandrasekhar, {\em The mathematical theory of black holes}, Oxford University Press, Oxford (1992).

\bibitem{lopez2020}M. Bouhmadi-López, S. Brahma, C.-Y. Chen, P. Chen, and D. Yeom, {\it A Consistent Model of Non-Singular Schwarzschild Black Hole in Loop Quantum Gravity and Its Quasinormal Modes}, {J. Cosmol. Astropart. Phys. {\bf 07}, 066 (2020)} {\tt [arXiv:2004.13061]}.

\bibitem{13} H. S. Liu, Z. F. Mai, Y. Z. Li and H. Lu, \textit{Quasi-topological Electromagnetism: Dark Energy, Dyonic Black Holes, Stable Photon Spheres and Hidden Electromagnetic Duality},
Sci. China Phys. Mech. Astron. \textbf{63}, 240411 (2020), { \tt arXiv:1907.10876 [hep-th]}.
\bibitem{17}A. Cisterna, G. Giribet, J. Oliva and K. Pallikaris, \textit{Quasitopological electromagnetism
and black holes}, Phys. Rev. D \textbf{101}, no.12, 124041 (2020),  {\tt arXiv:2004.05474 [hep-th]}.
\bibitem{18} M. D. Li, H. M.Wang and S. W.Wei, \textit{Triple points and novel phase transitions of dyonic
AdS black holes with quasi-topological electromagnetism}, { \tt arXiv:2201.09026 [gr-qc]}.
\bibitem{19}S. Dutta, A. Jain, R. Soni, \textit{Dyonic black hole and holography}. JHEP \textbf{1312}, 060 (2013), {\tt arXiv:1310.1748 [hep-th]}.
\bibitem{d}S. A. Hosseini Mansoori, \textit{Thermodynamic geometry of the novel 4-D Gauss–Bonnet AdS black
hole}, Phys. Dark Univ. \textbf{31} (2021) 100776, {\tt arXiv:2003.1338}.
\bibitem{014} D. V. Singh and S. Siwach, \textit{Thermodynamics and P-v criticality of Bardeen-AdS Black Hole in 4D Einstein-Gauss-Bonnet Gravity}, Phys. Lett. B \textbf{808}, 135658 (2020), {\tt arXiv:2003.11754 [gr-qc]}.
\bibitem{34}D. Glavan and C. Lin, \textit{Einstein-Gauss-Bonnet gravity in $4$-dimensional space-time}, Phys.
Rev. Lett. \textbf{124}, no. 8, 081301 (2020), { \tt arXiv:1905.03601}.
\bibitem{36}B. Eslam Panah, Kh. Jafarzade, S. H. Hendi, \textit{Charged 4D Einstein-Gauss-Bonnet-AdS Black
Holes: Shadow, Energy Emission, Deflection Angle and Heat Engine}, Nucl. Phys. B \textbf{961} 115269 (2020).
\bibitem{37}P. G. S. Fernandes, \textit{Charged Black Holes in AdS Spaces in 4D Einstein Gauss-Bonnet Gravity}, Phys. Lett. B \textbf{805} 135468 (2020).

\bibitem{new2023}D. J. Gogoi, Y. Sekhmani, D. Kalita, N. J. Gogoi, and J. Bora, Joule‐Thomson Expansion and Optical Behaviour of Reissner‐Nordstr\"om‐Anti‐de Sitter Black Holes in Rastall Gravity Surrounded by a Quintessence Field, Fortschritte Der Physik 2300010 (2023).

\bibitem{newref01}R.-G. Cai and K.-S. Soh, Topological Black Holes in the Dimensionally Continued Gravity, Phys. Rev. D {\bf 59}, 044013 (1999).
\bibitem{newref02}R.-G. Cai, Gauss-Bonnet Black Holes in AdS Spaces, Phys. Rev. D {\bf 65}, 084014 (2002).
\bibitem{newref03}P. G. S. Fernandes, Charged Black Holes in AdS Spaces in $4D$ Einstein Gauss-Bonnet Gravity, Physics Letters B {\bf 805}, 135468 (2020).

\bibitem{38}S. A. Hosseini Mansoori, \textit{Thermodynamic geometry of the novel 4-D Gauss–Bonnet AdS black hole}, Phys. Dark Univ. \textbf{31} (2021) 100776, {\tt arXiv:2003.1338}.
\bibitem{39}A.Belhaj and Y.Sekhmani,
\textit{Thermodynamics of Ay\'on-Beato\textendash{}Garc\'\i{}a\textendash{}AdS black holes in 4D Einstein\textendash{}Gauss\textendash{}Bonnet gravity}, Eur. Phys. J. Plus \textbf{137} (2022) no.2, 278.
\bibitem{41}
M.~Guo and P.~C.~Li,
\textit{Innermost stable circular orbit and shadow of the $4D$ Einstein\textendash{}Gauss\textendash{}Bonnet black hole},
Eur. Phys. J. C \textbf{80} (2020) no.6, 588, {\tt arXiv:2003.02523}.
\bibitem{42}
R.~A.~Konoplya and A.~F.~Zinhailo,
\textit{Quasinormal modes, stability and shadows of a black hole in the 4D Einstein\textendash{}Gauss\textendash{}Bonnet gravity},
Eur. Phys. J. C \textbf{80} (2020) no.11, 1049, {\tt arXiv:2003.01188}.
\bibitem{43}
R.~Kumar and S.~G.~Ghosh,
\textit{Rotating black holes in $4D$ Einstein-Gauss-Bonnet gravity and its shadow},
JCAP \textbf{07} (2020), 053
doi:10.1088/1475-7516/2020/07/053
{ \tt arXiv:2003.08927}.
\bibitem{44}
A.~Belhaj and Y.~Sekhmani,
\textit{Shadows of rotating quintessential black holes in Einstein\textendash{}Gauss\textendash{}Bonnet gravity with a cloud of strings},
Gen. Rel. Grav. \textbf{54} (2022) no.2, 17.
\bibitem{44a}
A.~Belhaj and Y.~Sekhmani,
\textit{Optical and thermodynamic behaviors of  Ay\'on-Beato\textendash{}Garc\'\i{}a\textendash{} black holes for 4D Einstein Gauss\textendash{}Bonnet gravity},
 Annals Phys. \textbf{441} (2022), 168863.
 \bibitem{BI} M. Born and L. Infeld, \textit{Foundations of the new field theory}, Proc. Roy. Soc. Lond.
A 144, no. 852, 425 (1934). doi:10.1098/rspa.1934.005
\bibitem{EH}W. Heisenberg and H. Euler, Z. Physik, 98, 714 (1936)
{\tt arXiv:physics/0605038}.
\bibitem{MM}I. Bandos, K. Lechner, D. Sorokin and P. K. Townsend, \textit{A non-linear duality-invariant conformal
extension of Maxwell’s equations}, Phys. Rev. D \textbf{102}, 121703 (2020), {\tt arXiv:2007.09092 [hep-th]}.
\bibitem{TC1} J. Oliva and S. Ray, \textit{A new cubic theory of gravity in five dimensions: Black hole,
Birkhoff’s theorem and C-function}, Class. Quant. Grav. \textbf{27}, 225002 (2010) doi:10.
1088/0264-9381/27/22/225002 {\tt arXiv:1003.4773 [gr-qc]}
\bibitem{TC2} R.C. Myers and B. Robinson, \textit{Black holes in quasi-topological gravity}, JHEP 1008,
\textbf{067} (2010) doi:10.1007/JHEP08(2010)067 {\tt arXiv:1003.5357 [gr-qc]}
\bibitem{TC3} M.H. Dehghani, A. Bazrafshan, R.B. Mann, M.R. Mehdizadeh, M. Ghanaatian and
M.H. Vahidinia, \textit{Black holes in quartic quasitopological gravity}, Phys. Rev. D \textbf{85}, 104009 (2012) doi:10.1103/PhysRevD.85.104009 {\tt arXiv:1109.4708 [hep-th]}
\bibitem{TC4} Y.Z. Li, H.S. Liu and H. L¨u, \textit{Quasi-topological Ricci polynomial gravities}, JHEP
1802, \textbf{166} (2018) doi:10.1007/JHEP02(2018)166 {\tt arXiv:1708.07198 [hep-th]}

\bibitem{new01}M. G\"urses, T. Ç. Sisman, and B. Tekin, {\it Is There a Novel Einstein–Gauss–Bonnet Theory in Four Dimensions?}, Eur. Phys. J. C {\bf 80}, 647 (2020).

\bibitem{new02}R. A. Hennigar, D. Kubiznak, R. B. Mann, and C. Pollack, {\it On Taking the $D \rightarrow 4$ Limit of Gauss-Bonnet Gravity: Theory and Solutions}, J. High Energ. Phys. {\bf 07}, 27 (2020).

\bibitem{new03}J. Arrechea, A. Delhom, and A. Jimenez-Cano, {\it Inconsistencies in Four-Dimensional Einstein-Gauss-Bonnet Gravity}, Chinese Phys. C {\bf 45}, 013107 (2021).

\bibitem{new04}H. Lu and Y. Pang, {\it Horndeski Gravity as $D \rightarrow 4$ Limit of Gauss-Bonnet}, Physics Letters B {\bf 809}, 135717 (2020).

\bibitem{new05}T. Kobayashi, {\it Effective Scalar-Tensor Description of Regularized Lovelock Gravity in Four Dimensions}, J. Cosmol. Astropart. Phys. {\bf 07}, 013 (2020).

\bibitem{new06}K. Jusufi, {\it Quasinormal Modes of Black Holes Surrounded by Dark Matter and Their Connection with the Shadow Radius}, Phys. Rev. D {\bf 101}, 084055 (2020).

\bibitem{new07}K. Jusufi, {\it Connection between the Shadow Radius and Quasinormal Modes in Rotating Spacetimes}, Phys. Rev. D {\bf 101}, 124063 (2020).

\bibitem{new08}B. Cuadros-Melgar, R. D. B. Fontana, and J. de Oliveira, {\it Analytical Correspondence between Shadow Radius and Black Hole Quasinormal Frequencies}, Physics Letters B {\bf 811}, 135966 (2020).

\end{thebibliography}
\end{document}